\documentclass[longauth]{aa}  
\usepackage{graphicx}
\usepackage{txfonts}
\usepackage{float}
\usepackage{times,epsfig,amsmath}
\usepackage{epstopdf}
\usepackage{pdflscape}
\usepackage{adjustbox}
\usepackage{amssymb, amsmath} 
\usepackage[usenames,dvipsnames]{color}
\usepackage{txfonts}
\usepackage{microtype}
\usepackage{url}
\usepackage[bookmarks, colorlinks, breaklinks]{hyperref}  
\hypersetup{linkcolor=blue,citecolor=blue,filecolor=blue,urlcolor=blue} 
\usepackage{booktabs}
\usepackage{subfig}
\usepackage{commath}
\usepackage{natbib}
\usepackage{amssymb}
\usepackage{pifont}
\bibpunct{(}{)}{;}{a}{}{,} 
\usepackage[bookmarks, colorlinks, breaklinks, ]{hyperref}  
\usepackage{booktabs}
\usepackage{soul}

\newfont{\gwpfont}{cmssq8 scaled 1000}
\newcommand{\rexcess}{{\gwpfont REXCESS}}

\begin{document} 
 \title{CHEX-MATE: the intracluster medium entropy distribution in the gravity-dominated regime}
 \author{G. Riva\inst{1,2} \and G. W. Pratt\inst{3} \and M. Rossetti\inst{1} \and I. Bartalucci\inst{1} \and S. T. Kay\inst{4} \and E. Rasia\inst{5,6} \and R. Gavazzi\inst{7,8} \and K. Umetsu\inst{9} \and M. Arnaud\inst{3} \and \\ M. Balboni\inst{1,10} \and A. Bonafede\inst{11,12} \and H. Bourdin\inst{13,14} \and S. De Grandi\inst{15} \and F. De Luca\inst{13,14} \and D. Eckert\inst{16} \and S. Ettori\inst{17,18} \and \\ M. Gaspari\inst{19} \and F. Gastaldello\inst{1} \and V. Ghirardini\inst{17} \and S. Ghizzardi\inst{1} \and M. Gitti\inst{11,12} \and L. Lovisari\inst{1,20} \and B. J. Maughan\inst{21} \and \\P. Mazzotta\inst{13,14} \and S. Molendi\inst{1} \and E. Pointecouteau\inst{22} \and J. Sayers\inst{23} \and M. Sereno\inst{17,18} \and I. Towler\inst{4}}
          
 \institute{INAF - Istituto di Astrofisica Spaziale e Fisica Cosmica di Milano, via A. Corti 12, 20133 Milano, Italy\\
              \email{giacomo.riva@inaf.it}     
         \and
             Universit\`a degli Studi di Milano, via G. Celoria 16, 20133 Milano, Italy
        \and
            Université Paris-Saclay, Université Paris Cité, CEA, CNRS, AIM, 91191, Gif-sur-Yvette, France
        \and    
            Jodrell Bank Centre for Astrophysics, Department of Physics and Astronomy, The University of Manchester, Oxford Road, Manchester M13 9PL, UK
        \and
            INAF, Osservatorio di Trieste, Via Tiepolo 11, 34131 Trieste, Italy
        \and 
            IFPU, Institute for Fundamental Physics of the Universe, Via Beirut 2, 34014 Trieste, Italy
        \and
            Laboratoire d’Astrophysique de Marseille, CNRS, Aix-Marseille Universit\'e, CNES, Marseille, France
        \and
            Institut d'Astrophysique de Paris, UMR 7095, CNRS, and Sorbonne Universit\'e, 98 bis boulevard Arago, 75014 Paris, France
        \and
            Academia Sinica Institute of Astronomy and Astrophysics (ASIAA), No. 1, Section 4, Roosevelt Road, Taipei 106216, Taiwan
        \and
            Dipartimento di Scienza e Alta Tecnologia, Università dell’Insubria, Via Valleggio 11, 22100 Como, Italy
        \and 
            Dipartimento di Fisica e Astronomia DIFA – Universit\`a di Bologna, via Gobetti 93/2, 40129 Bologna, Italy
        \and
            INAF – Istituto di Radioastronomia, via P. Gobetti 101, 40129 Bologna, Italy  
        \and
            Dipartimento di Fisica, Universit\`a di Roma ‘Tor Vergata’, Via della Ricerca Scientifica 1, 00133 Roma, Italy
        \and
            INFN, Sezione di Roma ‘Tor Vergata’, Via della Ricerca Scientifica 1, 00133 Roma, Italy
        \and 
            INAF – Osservatorio Astronomico di Brera, via E. Bianchi 46, 23807, Merate (LC), Italy
        \and
            Department of Astronomy, University of Geneva, ch. d’Écogia 16, CH-1290 Versoix Switzerland
        \and
            INAF, Osservatorio di Astrofisica e Scienza dello Spazio, Via Piero Gobetti 93/3, 40129 Bologna, Italy
        \and
            INFN, Sezione di Bologna, Viale Berti Pichat 6/2, 40127 Bologna, Italy
        \and 
            Department of Physics, Informatics and Mathematics, University of Modena and Reggio Emilia, 41125 Modena, Italy
        \and
            Center for Astrophysics | Harvard \& Smithsonian, 60 Garden St., Cambridge, MA 02138, USA
        \and
            HH Wills Physics Laboratory, University of Bristol, Tyndall Ave, Bristol, BS8 1TL, UK
        \and
            IRAP, CNRS, Université de Toulouse, CNES, UT3-UPS, Toulouse, France
        \and
            California Institute of Technology, 1200 East California Boulevard, Pasadena, California 91125, USA
        }

\date{Received 11 July 2024 / Accepted 08 October 2024}

\abstract{We characterise the intracluster gas entropy profiles of 32 very high mass ($M_{500} > 7.75 \times 10^{14}$ M$_{\odot}$) {\it Planck} SZ-detected galaxy clusters (HIGHMz), selected from the CHEX-MATE sample, allowing us to study the intracluster medium (ICM) entropy distribution in a regime where non-gravitational effects are expected to be minimised. Using XMM-{\it Newton} measurements, we determine the entropy profiles up to $\sim R_{500}$ for all objects.  We assess the relative role of gas density and temperature measurements on the uncertainty in entropy reconstruction, showing that in the outer regions the largest contribution comes from the temperature. The scaled profiles exhibit a large dispersion in the central regions, but converge rapidly to the value expected from simple gravitational collapse beyond the core regions. We quantify the correlation between the ICM morphological parameters and scaled entropy as a function of radius, showing that centrally-peaked objects have low central entropy, while morphologically disturbed objects have high central entropy. We compare the scaled HIGHMz entropy profiles to results from other observational samples, finding differences in normalisation which appear linked to the average mass of the samples in question. Combining HIGHMz with other samples, we find that a weaker mass dependence than self-similar in the scaling ($A_m \sim -0.25$) allows us to minimise the dispersion in the radial range $[0.3-0.8]\,R_{500}$ for clusters spanning over a decade in mass. The deviation from self-similar predictions is radially dependent and is more pronounced at small and intermediate radii than at $R_{500}$. We also investigate the distribution of central entropy $K_0$, finding no evidence for bimodality in the data, and outer slope $\alpha$, which peaks at $\alpha \sim 1.1$ with tails to both low and high $\alpha$ that correlate with dynamical state. Using weak lensing masses for half of the sample, we find an indication for a small suppression of the scatter ($\sim30\%$) beyond the core when using masses derived from $Y_X$ in the rescaling. Finally, we compare to recent cosmological numerical simulations from T{\tiny HE} T{\tiny HREE} H{\tiny UNDRED} and MACSIS, finding good agreement with the observational data in this mass regime. These results provide a robust observational benchmark in the gravity-dominated regime, and will serve as a future reference for samples at lower mass, higher redshifts, and for ongoing work using cosmological numerical simulations.}

\keywords{X-rays: galaxies: clusters -- galaxies: clusters: general -- galaxies : clusters : intracluster medium}
\maketitle

\section{Introduction}
Galaxy clusters are dark-matter dominated astrophysical objects with total masses\footnote{We define $R_{\Delta}$ as the radius inside which the cluster mass density is $\Delta$ times the critical density of the Universe $\rho_{\text{c}}(z)=3H_0^2 E(z)^2/8\pi G$, where $E(z)=\sqrt{\Omega_\text{m}(1+z)^3 + \Omega_{\Lambda}}$ and $z$ is the redshift; $M_{\Delta}$ is then the total mass within $R_{\Delta}$.} in the range $M_{500} \simeq 10^{14}-10^{15}$ M$_{\odot}$. Following the hierarchical scenario of structure formation, they reside at the nodes of the cosmic web and grow through accretion of matter along the filaments of the large-scale structure and through episodic mergers of small mass systems. Falling into the potential well of the dark matter, the intracluster medium (ICM) is heated to X-ray emitting temperatures ($\sim 10^7 - 10^8$ K) by shocks and compression. Given the scale-free nature of gravity, the resulting X-ray cluster population is expected to be self-similar and for tight scaling relations to exist between the ICM observables and cluster mass and redshift (\citeauthor{kaiser86} \citeyear{kaiser86}, \citeauthor{bryan98} \citeyear{bryan98}). However, second-order effects, mainly linked to feedback from active galactic nuclei (AGN) and radiative cooling of the gas, act to modify the properties of the ICM and induce some degrees of departure from self-similar predictions (see, e.g., \citeauthor{voi05_rev} \citeyear{voi05_rev}; \citeauthor{pratt09} \citeyear{pratt09}; \citeauthor{pratt10} \citeyear{pratt10};
\citeauthor{borgani10} \citeyear{borgani10}; \citeauthor{giodini13} \citeyear{giodini13}; 
\citeauthor{gaspari20} \citeyear{gaspari20}; \citeauthor{lovisari22} \citeyear{lovisari22}). 

In recent years, spatially resolved observations have allowed for a more detailed examination of the impact of non-gravitational processes on the ICM. In particular, the study of the thermodynamic properties of the ICM has been of great interest, since it allows us to obtain useful insights on the history and level of the energy deposited in the ICM through feedback processes. Gas entropy, defined as $K = T/n_e^{2/3}$ (\citeauthor{ponman99} \citeyear{ponman99}; \citeauthor{lloyd-davies00} \citeyear{lloyd-davies00}), where $T$ is the gas temperature and $n_e$ the electron density\footnote{We note that the adopted definition of gas entropy is a convention commonly used in X-ray astronomy. It is linked to the classic thermodynamic specific entropy by a logarithm and a constant.}, plays a key role in this context, since it both determines the structure of the ICM and provides a record of the processes (both gravitational and non gravitational) that influence the properties of the ICM. Entropy is generated during the hierarchical assembly process and then is modified by any other process that can change the physical characteristics of the gas (e.g., AGN heating, cooling and star formation). For these reasons, it is the ideal tool for investigating the thermodynamic history of the cluster population \citep[see][for a review]{voi05_rev}

In a stable ICM in hydrostatic equilibrium, stratification naturally results in a radially increasing entropy profile. \citet{tozzi01} used analytical modelling to study the characteristic radial distribution of the entropy, where outside the core region, it was shown to steadily increase with radius, following a power law with a slope of $K \propto R^{1.1}$. This conclusion was reinforced by the non-radiative (gravity-only) simulations of \citet{voit+05}, where, again outside the core, the entropy slope was found to be remarkably stable out to $\sim2~R_{200}$, independent of system mass. These predictions are regarded as robust and depend little on the adopted numerical approach outside of the central region (\citeauthor{mitchell09} \citeyear{mitchell09}; \citeauthor{gaspari12} \citeyear{gaspari12}).

Early observations, however, showed that the entropy profiles of lower-mass systems exceeded that expected from gravity alone \citep{lloyd-davies00}. Subsequent observations have refined this picture. The mass dependence of the entropy excess has been confirmed, and has been shown to extend to larger radius in lower-mass systems \citep{pratt10}. As a result, lower mass systems generally exhibit a shallower entropy slope, with significantly increased core entropy relative to the expectation from simple gravitational collapse. Moving outwards, the observed profiles converge towards the non-radiative prediction, but the convergence radius is larger for less massive systems. This contributes to the observed large scatter in the central regions (e.g. \citeauthor{cavagnolo09} \citeyear{cavagnolo09}; \citeauthor{pratt10} \citeyear{pratt10}; \citeauthor{ghirardini19} \citeyear{ghirardini19}). The now-standard explanation for this behaviour is that it is due to the combined effect of non-gravitational processes. These include, but are not limited to: radiative cooling of the central gas, which removes material from the hot phase, leaving only gas on a higher adiabat; stochastic feedback from the central AGN over time, which injects energy into the ICM and raises its entropy;  feedback from supernovae (SNe) in the cluster galaxies, which also inject energy into the ICM; conduction, which smooths out temperature (and therefore entropy) gradients. While numerical simulations that include these processes can go some way to explaining the observed entropy distributions, they still struggle to reproduce the observed population behaviour \citep[see e.g.][]{nagai07,barnes17,altamura23,oppenheimer21,kayp22}.

Interpretation of the observed properties of the ICM entropy is complicated by the clear dependence of the entropy distribution on the morphological properties of the cluster gas (\citeauthor{pratt10} \citeyear{pratt10}). Relaxed systems generally possess cool cores, and such objects exhibit an entropy distribution that closely follows the $K \propto R^{1.1}$ distribution down to very small radii. In contrast, disturbed systems always exhibit high core entropy \citep[e.g.][]{pratt10,babazaki18}. However, the central surface brightness of morphologically disturbed systems is always flatter than that of cool core systems of similar mass. This means that at a given mass, to obtain a spectrum with a given number of counts or signal-to-noise ratio, a larger central region must be used for morphologically disturbed systems than for cool core systems. The resulting entropy profiles therefore have poorer angular resolution in the centre, further complicating the interpretation.

In this context, it is interesting to revisit the entropy distribution of the highest-mass systems. In such objects, gravity is the dominant entropy generation mechanism, and non-gravitational effects should be minimised. As such, they should provide a robust baseline for theoretical works. In addition, such systems are highly luminous in X-rays, allowing measurement of the entropy both deep into the core and out to large radius. Here we use a subset of 32 very high mass ($M_{500} > 7.75 \times 10^{14}$ M$_{\odot}$) clusters, which we have dubbed HIGHMz, selected from the "Cluster HEritage project with XMM-\textit{Newton}: Mass Assembly and Thermodynamics at the Endpoint of structure formation" (CHEX-MATE; \citeauthor{chexmate21} \citeyear{chexmate21}), a multiyear Heritage program to observe 118 {\it Planck} SZ-selected clusters with XMM-\textit{Newton}. CHEX-MATE observations are tailored for deriving the total masses, through the hydrostatic equilibrium equation, and characterising the thermodynamic properties of the entire sample with good precision within $R_{500}$. In this context, the very high mass of the systems in HIGHMz allows us to further investigate the ICM entropy and its radial distribution in the gravity-dominated regime, and to compare our results with recent large-volume numerical simulations including non-gravitational sub-grid physics. The results from this work will thus set an anchor for future studies at lower mass scales and higher redshifts.

The paper is organised as follows. In Sect. \ref{samples_sec} we present the HIGHMz sample, together with additional samples (both observational and simulated) used for comparison; in Sect. \ref{data processing} we describe our methods for the data processing and analysis; Section \ref{results} is dedicated to the results of this work: here we present the measured entropy profiles and compare them to other samples. Finally, we discuss our findings and provide a conclusion to the work in Sects. \ref{discussion} and \ref{summary}. Throughout the paper, we assume a flat $\Lambda$ cold dark matter (CDM) cosmology with $H_0 = 70$ km s$^{-1}$ Mpc$^{-1}$ (i.e. $h_{70}=1$), $\Omega_{\text{m}} = 0.3$ and $\Omega_{\Lambda} = 0.7$. The solar abundance table is set to \citet{asplund09}, while all the quoted errors hereafter are at the $1\sigma$ confidence level. Whenever we use the notation $M_{500}^{Y_{\rm SZ}}$ in this paper, we refer to the masses derived from the \textit{Planck} SZ \citep{sunyaev72} signal with the method described in \citet{planck16}, using the MMF3 algorithm \citep{melin06}, as discussed in \citet{chexmate21}.

\section{Samples}
\label{samples_sec}

\subsection{HIGHMz: high mass clusters}
For the present work, we have selected the most massive systems of the CHEX-MATE \citep{chexmate21} project. This sample, which we have dubbed HIGHMz, comprises 35 clusters at redshift $z>0.2$ and with $M_{500}^{Y_{\rm SZ}} > 7.75 \times 10^{14}$ M$_{\odot}$.  
\begin{figure}
\centering
            \includegraphics[width=0.47\textwidth]{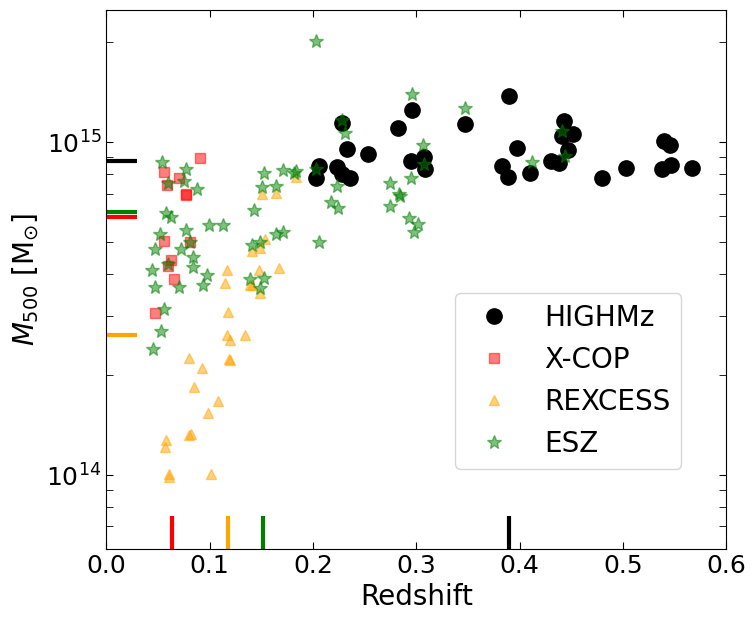}
            \caption{\footnotesize Distribution of HIGHMz (black dots) clusters in the mass-redshift plane. X-COP (red squares), \rexcess\ (orange triangles) and ESZ (green stars) clusters are also plotted for comparison. Vertical  and horizontal coloured lines mark the median redshift and mass, respectively, of the four samples.}
            \label{samples}
\end{figure}

\begin{table*}
\caption{\footnotesize HIGHMz sample: cluster properties and information on the used XMM-\textit{Newton} observations.}
\centering 
\small
\begin{tabular}{ccccccccc}        
\toprule   
\toprule
Cluster & Obs. IDs & Redshift & $M_{500}^{Y_{\rm SZ}}$  & $R_{500}^{Y_{\rm SZ}}$ & $c$ &$w$ & Dyn. state & $S/B<0.2$\\
&&& $10^{14}$ M$_{\odot}$ & Mpc & &  & &\\
\midrule  
PSZ2G004.45$-$19.55 &  0656201001 / 0827050301 & 0.540 & 10.090 & 1.254 & 0.37 & 0.006 & ...&...\\
PSZ2G008.94$-$81.22 &  0743850101 & 0.307 & 8.989 & 1.321 & 0.19 & 0.053 & D &...\\
PSZ2G044.77$-$51.30 &  0693661901 / 0827060401 & 0.503 & 8.359 & 1.195 & 0.34 & 0.006 & ...&\checkmark\\
PSZ2G046.10$+$27.18 &  0827060201 / 0723160601 & 0.389 & 7.840 & 1.223 & 0.15 & 0.022 & ...&...\\
PSZ2G056.93$-$55.08 &  0503490201 & 0.447 & 9.491 & 1.275 & 0.17 & 0.025 & ...&...\\
PSZ2G057.25$-$45.34 &  0693010601 & 0.397 & 9.624 & 1.306 & 0.46 & 0.008 & ...&\checkmark\\
PSZ2G072.62$+$41.46 &  0605000501 & 0.228 & 11.426 & 1.473 & 0.32 & 0.005 & R &\checkmark\\
PSZ2G073.97$-$27.82 &  0111270101 & 0.233 & 9.516 & 1.383 & 0.44 & 0.011 & ...&...\\
PSZ2G092.71$+$73.46 &  0084230901 & 0.228 & 8.003 & 1.308 & 0.29 & 0.005 & R &...\\
PSZ2G111.61$-$45.71 &  0111000101 / 0827061301 & 0.546 & 8.499 & 1.182 & 0.23 & 0.006 & ...&...\\
PSZ2G155.27$-$68.42 &  0693662801 / 0827060801 / 0920890401 & 0.567 & 8.364 & 1.166 & 0.25 & 0.027 & ...& ...\\
PSZ2G159.91$-$73.50 &  0084230301 & 0.206 & 8.464 & 1.343 & 0.31 & 0.019 & ...&...\\
PSZ2G186.37$+$37.26 &  0827041001 & 0.282 & 10.998 & 1.426 & 0.32 & 0.007 & ...&\checkmark\\
PSZ2G195.75$-$24.32 &  0201510101 & 0.203 & 7.800 & 1.309 & 0.20 & 0.011 & ...&\checkmark\\
PSZ2G201.50$-$27.31 &  0205670101 / 0827061801 & 0.538 & 8.304 & 1.176 & 0.32 & 0.010 & ...&...\\
PSZ2G205.93$-$39.46 &  0827011501 & 0.443 & 11.542 & 1.363 & 0.38 & 0.006 & ...&...\\
PSZ2G210.64$+$17.09 &  0658200501 / 0827360301 & 0.480 & 7.790 & 1.178 & 0.19 & 0.019 & ...&\checkmark\\
PSZ2G216.62$+$47.00 &  0827340901 & 0.383 & 8.469 & 1.258 & 0.40 & 0.024 & ...& ...\\
PSZ2G228.16$+$75.20 &  0693661701 / 0827341301 & 0.545 & 9.790 & 1.239 & 0.21 & 0.030 & D & \checkmark\\
PSZ2G239.27$-$26.01 &  0827010401 & 0.430 & 8.772 & 1.250 & 0.24 & 0.030 & ...& \checkmark\\
PSZ2G243.15$-$73.84 &  0827011301 & 0.410 & 8.086 & 1.226 & 0.16 & 0.022 & ...& ...\\
PSZ2G262.27$-$35.38 &  0692934301 & 0.295 & 8.759 & 1.315 & 0.14 & 0.035 & D & ...\\
PSZ2G266.04$-$21.25 &  0112980201 & 0.296 & 12.470 & 1.479 & 0.27 & 0.013 & ...& ...\\
PSZ2G277.76$-$51.74 &  0674380301 & 0.438 & 8.650 & 1.240 & 0.14 & 0.037 & D & ...\\
PSZ2G278.58$+$39.16 &  0042341001 / 0827051201 & 0.308 & 8.290 & 1.285 & 0.32 & 0.037 & D & ...\\
PSZ2G284.41$+$52.45 &  0762070101 & 0.441 & 10.400 & 1.317 & 0.40 & 0.002 & R & ...\\
PSZ2G286.98$+$32.90 &  0656201201 / 0827341401 & 0.390 & 13.742 & 1.474 & 0.22 & 0.017 & ...& \checkmark\\
PSZ2G324.04$+$48.79 &  0112960101 & 0.452 & 10.579 & 1.319 & 0.60 & 0.002 & R & ...\\
PSZ2G340.36$+$60.58 &  0551830101 & 0.253 & 9.199 & 1.358 & 0.62 & 0.002 & R & \checkmark\\
PSZ2G340.94$+$35.07 &  0827311201 & 0.236 & 7.795 & 1.293 & 0.57 & 0.007 & ...& \checkmark\\
PSZ2G346.61$+$35.06 &  0827311401 & 0.223 & 8.409 & 1.332 & 0.14 & 0.060 & D & \checkmark\\
PSZ2G349.46$-$59.95 &  0504630101 & 0.347 & 11.359 & 1.407 & 0.44 & 0.005 & R &\checkmark\\
\bottomrule                                  
\end{tabular}
\tablefoot{In the Table, we list the PSZ2 name of the clusters, the ID of the used XMM-\textit{Newton} observations, the redshift, the nominal integrated mass estimated from \textit{Planck} SZE data \citep{planck16}, and two morphological indicators, $c$ and $w$, as measured by \citet{campitiello22}. Errors on $c$ and $w$ can be found in Table A.1 of \citet{campitiello22}. We also indicate the most disturbed (D) and relaxed (R) clusters, classified according to their $w$ values, which are used in Sect. \ref{sect_corr_c_w}. Finally, we report whether a cluster has an external low source-to-background ($S/B < 0.2$) temperature measurement. A gallery of the cluster images is shown in Fig. 6 of \citet{chexmate21}}. 
\label{table:1}
\end{table*}

We excluded from the present analysis three galaxy clusters, namely PSZ2G107.10+65.32 (a.k.a. A1758N), PSZ2G225.93-19.99 (a.k.a. MACS J0600-20) and PSZ2G339.63-69.34 (a.k.a. Phoenix). The former two are double clusters and do not allow a simple radial analysis. The latter has a strong central AGN, whose emission, given the size of XMM-\textit{Newton}'s PSF, has an impact also on regions beyond the core. Since including these three complex clusters may introduce systematics in our results, we preferred to keep them aside and to include them subsequently in the final study on the full CHEX-MATE sample, adopting an \textit{ad-hoc} analysis. The final list of the 32 analysed HIGHMz clusters is presented in Table \ref{table:1}, together with information about cluster masses, redshifts and indicators for their morphological state (light concentration, $c$; centroid shift, $w$), as measured by \citet{campitiello22}. High values of $c$ and low values of $w$ are commonly measured for clusters in a dynamically relaxed state, while low $c$ and high $w$ are commonly associated with disturbed systems \citep[e.g.][]{santos08,hudson10,lovisari17,campitiello22}. The cluster masses and redshifts are also plotted in Fig. \ref{samples}, as black dots.

Given its selection in mass and redshift, HIGHMz allows us to address multiple and interesting aims. In particular, it offers the opportunity to study the cluster entropy in a regime where gravitational processes are considered dominant, thus setting a valuable baseline for future studies at lower mass scales, and to compare with local samples available in the literature, thus investigating potential hints of variation of the gas entropy profiles with mass and/or redshift.


\subsection{Comparison samples: \rexcess, ESZ and X-COP}
\label{sample_presentation}

We will compare with three cluster samples available in the literature, namely \rexcess\ \citep{bohringer07}, ESZ \citep{planck11_ESZ62} and X-COP \citep{eckert17}. All of these samples have been observed with XMM-\textit{Newton} and thus allow direct comparison with CHEX-MATE. The four samples together allow investigation of the gas entropy in clusters spanning a wide range both in mass and redshift, thus giving us the possibility to test predictions from self-similar scenario. \rexcess, ESZ and X-COP are extensively described in the referenced papers; however we provide here below a summary of their main features:
\begin{itemize}
    \item \rexcess\ consists of 31 nearby ($z < 0.2$) galaxy clusters with temperatures in the range $2-9$ keV, drawn from the REFLEX catalogue \citep{bohringer04}. Therefore, it provides a census of the local X-ray selected cluster population. \rexcess\ clusters were selected in X-ray luminosity only, with no bias towards any particular morphological type;
    \item the original ESZ sample comprises 189 clusters that were identified via their SZ effect in the first all-sky coverage by the \textit{Planck} satellite \citep{planck11, planck11_ESZ}. After cross-correlating with the MCXC catalogue \citep{piffaretti11} and checking the quality of XMM-\textit{Newton} archive data, the suitable sample for the X-ray analysis was reduced to 62 clusters. This final ESZ sample consists of clusters at $z < 0.5$, spanning one decade in mass;
    \item X-COP is a set of 12 massive and local ($z < 0.1$) clusters selected from the \textit{Planck} all-sky survey of SZ sources (\citeauthor{planck14} \citeyear{planck14}; \citeauthor{planck16} \citeyear{planck16}). The aim of the X-COP project was to advance the knowledge of the physical conditions in the cluster outskirts, by combining high signal-to-noise SZ data with good X-ray data coverage at and beyond $\sim R_{500}$. The thermodynamic profiles of X-COP clusters from joint XMM-\emph{Newton} and \emph{Planck} data were published in \citet{ghirardini19}. The entropy profiles used here slightly differ from the ones presented in \citet{ghirardini19} as they are the result of a joint non-parametric reconstruction of the deprojected thermodynamic profiles, whereby the 3D temperature profile is described as a linear combination of a large number of log-normal functions sampling the radial range of interest. For more details on the deprojection technique, we refer the reader to \citet{eckert22}.

\end{itemize}

Masses and redshifts of \rexcess, ESZ and X-COP samples are shown in Fig. \ref{samples} and compared to HIGHMz. In particular, we plot i) $M_{500}^{Y_{\rm SZ}}$ masses for HIGHMz; ii) masses measured using the $M_{500}-Y_X$ scaling \citep{kravtsov06}, as calibrated in \citet{arnaud10}, for \rexcess\ and ESZ; and iii) hydrostatic equilibrium masses for X-COP, as computed in \citet{eckert22}. The HIGHMz sample has both a higher average mass and a higher average redshift than the other samples. We note that ESZ has 10, 8 and 7 clusters in common with HIGHMz, X-COP and \rexcess, respectively, while the other three samples do not have clusters in common with each others. We checked that the entropy profiles of these common clusters, as measured for the different samples, are in agreement within the errors.


\subsection{Simulated samples: MACSIS and The300}
\label{sim_presentation}

We also compare our results with recent hydrodynamic simulations that are able to reproduce a sufficient number of massive clusters in the mass range of interest. In particular, we consider two simulated datasets, taken from the parent MACSIS \citep{barnes17} and T{\tiny HE} T{\tiny HREE} H{\tiny UNDRED} (hereafter The300; \citeauthor{cui18} \citeyear{cui18}) projects. Briefly:
\begin{itemize}
    \item MACSIS is a sample of 390 massive clusters selected from a large-volume ($3.2\,{\rm Gpc}$ box size) dark matter simulation, which were then re-simulated with full gas physics. MACSIS extends the BAHAMAS ($400\,h^{-1}\,{\rm Mpc}$ box size) simulation \citep{mccarthy17} to the most massive clusters expected to form  in a $\Lambda$CDM cosmology. These were both run with a heavily modified version of the {\small GADGET3} TreePM-SPH code, last described in \cite{springel05}. The code incorporates radiative cooling on an element-by-element basis; heating and cooling from UV/X-ray and cosmic microwave backgrounds; star formation; stellar evolution and enrichment from AGB stars and Type~Ia/II SNe; kinetic stellar feedback; super-massive black hole growth and thermal AGN feedback. MACSIS was run with the same sub-grid parameter choices   as BAHAMAS, calibrated to match the low redshift galaxy stellar mass function and cluster gas fractions (see \citealt{mccarthy17} for further details of the baryonic physics model and calibration process);
    \item The300 ({\small GADGET-X} version) consists of the 324 most massive clusters identified at $z = 0$ within the dark matter only MultiDark simulation \citep{klypin16} of size $1 h^{-1}$ Gpc and resimulated accounting for baryons. The hydrodynamic code, implementing the SPH description as in \cite{beck.etal.2016}, is similar to what was used in \cite{rasia15} and includes metal-dependent radiative gas cooling, star formation, stellar feedback by asymptotic-giant-branch stars, Type Ia and core-collapse SNe, supermassive black hole growth, and AGN feedback.
\end{itemize}
Given their properties, both datasets are therefore ideal samples to test the comparison with the massive clusters in HIGHMz. 

\begin{figure}
            \centering\includegraphics[width=0.38\textwidth]{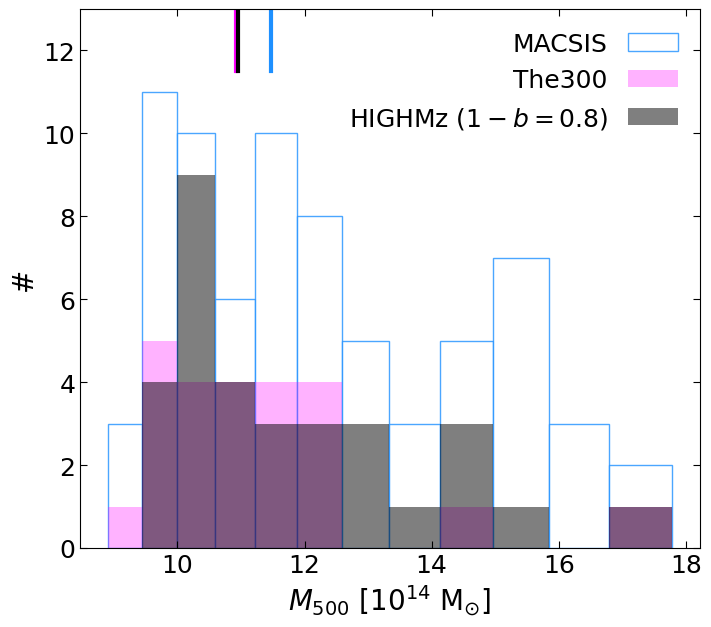}\\
            \hspace{-0.2cm}\includegraphics[width=0.41\textwidth]{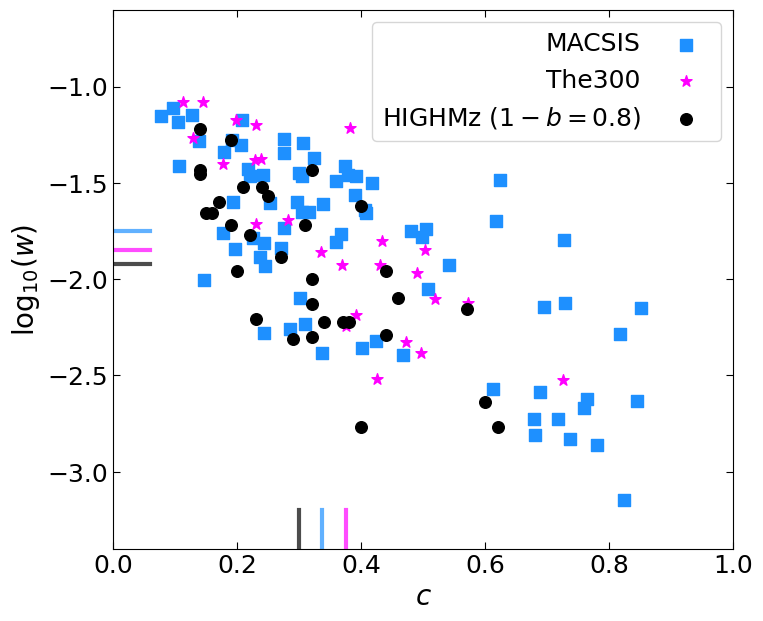}
            \caption{\footnotesize Properties of MACSIS (light blue) and The300 (violet) clusters, in comparison to HIGHMz (black). Top panel: distribution  of the cluster masses. Bottom panel: distribution in the concentration-centroid shift plane. In both panels, coloured lines show the median value of the corresponding axis for the three samples. MACSIS and The300 clusters are all at redshifts $z = 0.34$ and $z=0.333$, respectively.}
            \label{samples_sim}
\end{figure}

Following \citet{planckXX14}, we assumed that the mass measurements $M_{500}^{Y_{\rm SZ}}$ show an average hydrostatic bias of $20\%$ (i.e. $1-b = 0.8$). To ensure a fair comparison with simulations, which measure the true masses of the reproduced clusters, we rescaled HIGHMz masses accordingly, whose `true' values were considered to be $M_{500}^{Y_{\rm{SZ},corr.}} = M_{500}^{Y_{\rm SZ}}/0.8$. In order to define the most suitable simulated dataset for comparison with HIGHMz, we thus imposed MACSIS and The300 cluster masses to be $\gtrsim M_{500}^{Y_{\rm {SZ},corr.}}$, as also previously done in \citet{bartalucci+23}. The resulting sample contains 25 clusters from The300 and 75 from MACSIS, taken from one snapshot corresponding to redshift $z = 0.333$ (The300) and $z=0.34$ (MACSIS), the closest one for each sample to the median redshift of HIGHMz ($z = 0.39$). Similarly to observations, we also use information on morphological indicators (in particular, $c$ and $w$) for The300 and MACSIS, computed as detailed in \citet{campitiello22} and \citet{towler23}, respectively. Figure \ref{samples_sim} shows the comparison of both cluster masses and morphological indicators for the three samples, while the complete list of the selected simulated datasets is reported in Tables \ref{table:the300} and \ref{table:MACSIS}. Both simulation datasets adopt a flat $\Lambda$CDM cosmology with $h=0.678$ km s$^{-1}$ Mpc$^{-1}$, $\Omega_m = 0.307$ and $\Omega_\Lambda=0.693$ \citep{planck_xiii_16}. We corrected for the difference in the adopted cosmology when comparing to observations (Eq. \ref{eq_k500}).


\section{Analysis procedures}
\label{data processing}

The X-ray data have been reduced and analysed using the CHEX-MATE pipeline, which is extensively described in \citet{bartalucci+23} and \citet{rossetti24}. In the following, we report the main steps of our procedure and present the emission measure and gas temperature profiles of HIGHMz clusters. Finally, we also summarise our deprojection techniques to build the three-dimensional entropy profiles and provide information on the adopted self-similar scaling.

\subsection{Data reduction}

The HIGHMz clusters were observed with the European Photon Imaging Camera (EPIC, \citeauthor{turner01} \citeyear{turner01}; \citeauthor{struder01} \citeyear{struder01}) on board XMM-\textit{Newton}. The datasets were reprocessed using the Extended-Science Analysis System (ESAS, \citeauthor{snowden08} \citeyear{snowden08}) embedded in SAS version 16.1 (see \citealt{rossetti24} for details of this choice). We removed flare events by using the tools \textit{mos-filter} and \textit{pn-filter}, by extracting the light curves in the $[2.5-8.5]$ keV energy range and excising time intervals with count rates exceeding $3\sigma$ times the mean. Point sources were filtered from the analysis following the scheme detailed in Sect. 2.2.3 of \citet{ghirardini19} and summarised in Sect. 3.1.1 of \citet{bartalucci+23}.

For each cluster, we used the XMM-\textit{Newton} observations listed in Table \ref{table:1}. When multiple pointings were available, we combined them to increase the available count statistics, both for the image and the spectral analysis, as described in \citet{bartalucci+23} and specified in Sect. 3.3 of \citet{rossetti24}.

\subsection{Projected profiles}
\label{sect_2D}
\subsubsection{Emission measure}

The procedure we followed to derive HIGHMz emission measure profiles is outlined in detail in Sect. 3 of \citet{bartalucci+23}. Briefly, we first produced the EPIC image for each cluster in the energy band $0.7-1.2$ keV, together with exposure and background maps, merging together those of the three XMM-\textit{Newton}'s cameras. From the combination of these images, we then extracted both azimuthal mean and median background-subtracted and exposure-corrected surface brightness (SB) profiles, from the coordinates of the X-ray peaks. The use of the azimuthal mean allows inspection of the central regions of a cluster with higher resolution\footnote{To produce azimuthal median profiles, one first needs to produce an image of the cluster, with a Voronoi tessellation that guarantees a certain minimum number of counts per bin (see the discussion in Sect. 3.3 of \citealp{eckert15}). This necessarily results in a lower resolution in the central regions with respect to mean profiles, which are obtained by averaging the photons collected in the radial bin of interest and can thus fully exploit the native resolution of the instrument.} (e.g. \citeauthor{pratt22} \citeyear{pratt22}); conversely, the technique of computing azimuthal median SB profiles was introduced by \citet{eckert15} to limit the impact of sub-clumps and sub-structures too faint to be identified and masked (e.g. \citeauthor{roncarelli13} \citeyear{roncarelli13}; \citeauthor{zhuravleva13} \citeyear{zhuravleva13}). As specified later, we will make use of both these products in the computation of the ICM entropy. Finally, the SB profiles were converted to emission measures following Eq. 1 of \citet{arnaud02}. Emission measures are directly related to the gas density of the cluster, being the integral of the density squared along the line of sight. They are therefore necessary ingredients to recover the 3D density profiles.

\begin{figure*}
\centering
            \includegraphics[width=0.97\textwidth]{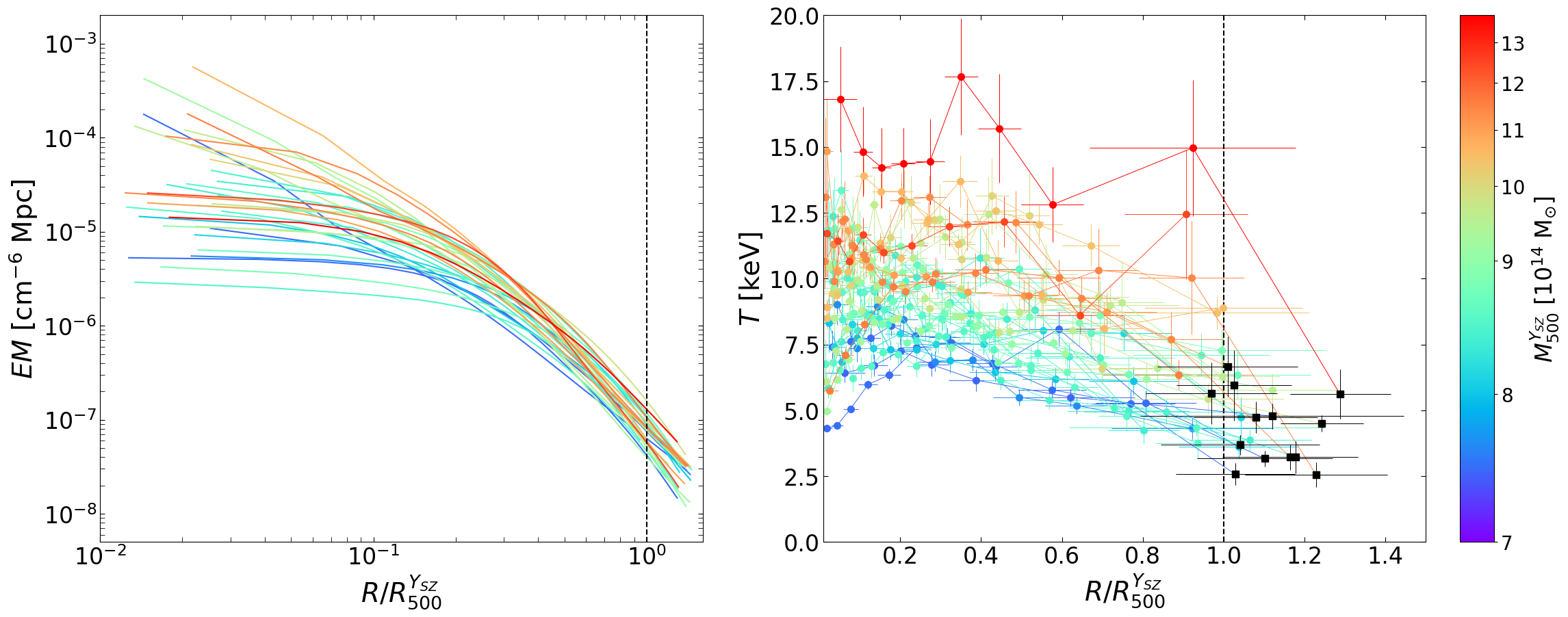}
            \caption{\footnotesize Azimuthal median emission measure (left) and projected temperature (right) profiles of clusters in HIGHMz. The profiles are colour coded according to cluster masses ($M_{500}^{Y_{\rm SZ}}$; Table \ref{table:1}). Temperature measurements with $S/B<0.2$ are marked in black. }
            \label{2D_prof}
\end{figure*}

The azimuthal median emission measure profiles of the HIGHMz clusters are presented in Fig. \ref{2D_prof} (left) and colour coded according to their masses $M_{500}^{Y_{\rm SZ}}$, with blue to red indicating from less to more massive clusters. As highlighted by \citet{bartalucci+23} for the entire CHEX-MATE sample, in the central regions of HIGHMz clusters we find considerable dispersion, with cool core clusters exhibiting steeper profiles, while moving towards the outskirts the profiles appear more self similar.

\subsubsection{Gas temperature}
\label{T_2D}

The steps of the CHEX-MATE pipeline built for the spectral extraction and fitting, which allowed us to measure the gas temperature profiles of HIGHMz clusters, are detailed in Sects. 3 and 4 of \citet{rossetti24}. The pipeline combines the best practices developed during previous projects (e.g. \citeauthor{pratt07} \citeyear{pratt07}; \citeauthor{planck11_ESZ62} \citeyear{planck11_ESZ62}; \citeauthor{bartalucci18} \citeyear{bartalucci18}; \citeauthor{ghirardini19} \citeyear{ghirardini19}), with the introduction of some novelties. These include, for example, the construction of a physical model for the particle background for all EPIC detectors and the application of a Bayesian MCMC framework, allowing the propagation of the uncertainties on the background parameters up to the final results of the spectral analysis.

An important point of strength of the CHEX-MATE pipeline is the careful analysis of the underlying systematics. Based on the study of a sample of 30 clusters, selected to be representative of the entire CHEX-MATE population, \citet{rossetti24} showed that we can obtain reliable temperature estimates at least up to regions where the source intensity is larger than $20\%$ of the background. Below this value (i.e. where $S/B<0.2$) temperature measurements can be biased low and the exclusion of these bins resulted in a better reconstruction of the mean temperature profile in the outer regions \citep{rossetti24}. 

In the spectral fitting, we fixed the redshift and Galactic absorption, but the temperature, metallicity, and normalisation of the cluster spectra were left free to vary, as described in \citet{rossetti24}.  The measured temperature profiles of HIGHMz clusters are presented in Fig. \ref{2D_prof} (right). Temperatures range from $\sim 5$ to $\sim 15$ keV, reflecting the high masses of the clusters in the sample. The adopted colour coding is the same as for the emission measures and highlights the dependence on cluster mass. Temperature measurements where $S/B < 0.2$ are also shown in the figure and marked in black. Almost all of these are located beyond $R_{500}$, while CHEX-MATE observations are tailored to measure temperatures within this radius. In about four cases, these measurements exhibit very low temperature values. Including these external bins in the deprojection can introduce a bias on the final products, as discussed later.

\subsection{Measuring the ICM entropy}
\label{sect_deproj}

Following the astrophysical convention, we construct the ICM entropy as $K = T/n_e^{2/3}$, using the three-dimensional profiles of gas density and temperature. The deprojection of both the emission measure and temperature profiles presented in Sect. \ref{sect_2D} is therefore needed in advance.

The emission measures were deprojected and PSF-corrected using the non-parametric method described in \citet{croston06}. As also detailed in \citet{croston08}, the emission measure was converted to gas density by calculating a global conversion factor for each profile in Xspec v. 12.13 \citep{arnaud96}, using the spectroscopic temperature in the $[0.15-1]~R_{500}^{Y_{\rm SZ}}$ aperture. A secondary correction factor, taking into account radial variations of temperature and abundance, was also included to give the final gas density profiles. This was obtained from analytical fits to the projected quantities \citep[see][]{pratt03}. Both azimuthal mean and median density profiles were produced and used to calculate the ICM entropy. We will consider the entropy profiles derived from azimuthal median densities as our reference throughout the paper. However, azimuthal mean densities will be used in Sect. \ref{other_samples}, for comparison with other samples.

To deproject and PSF-correct the 2D temperature profiles of HIGHMz clusters, we adopted the non-parametric-like technique described in \citet{dem10} and \citet{bartalucci18}. Briefly, we assumed that the 3D temperature profiles can be described by a parametric model, adapted from \citet{vikhlinin06}, that is convolved with a response matrix which simultaneously takes into account projection and PSF redistribution. The projection procedure additionally took into account the bias introduced by fitting isothermal models to multi-temperature plasma \citep{mazzotta04,vik06w}. The final 3D temperature profile is then estimated at the weighted radii corresponding to the 2D annular binning scheme. Low and external temperature measurements with $S/B < 0.2$, often accompanied with small statistical errors, have an impact on the deprojection. This is particularly true for the cluster outskirts, but also for the internal regions. We will show in Sect. \ref{sec_ent_results}, and discuss in more detail in Sect. \ref{sect_flattening}, the impact of including these low-$S/B$ measurements on the shape of the entropy profiles. Unless otherwise stated, in the following sections we will consider the entropy profiles constructed from temperature measurements with $S/B>0.2$ as our reference.

\subsection{Scaling and self-similar predictions}
Among the products of our deprojection pipeline are also integrated masses within $R_{500}$, that we adopted for rescaling the entropy profiles. In particular, temperature, density, and their gradients were used to compute both hydrostatic masses (hereafter $M_{500}^{\rm HE}$) and the quantity $Y_X=M_{\rm gas} \times T$ \citep{kravtsov06}, which was used as a proxy for the total mass (hereafter $M_{500}^{Y_{\rm X}}$), using the scaling calibrated in \citet{arnaud10}. Hydrostatic masses were measured using azimuthal median densities and adopted in the rescaling of the ICM entropy in comparison with X-COP clusters only (Sect. \ref{other_samples}); masses from $Y_X$ were computed both from median and mean densities and used in the following sections depending on the relevant study. 

According to the self-similar scenario, the physical properties of galaxy clusters should coincide, once they are correctly scaled for cluster mass and redshift. For the ICM entropy, the predicted dependence on mass and redshift from pure gravitational collapse is $K \propto M^{2/3}~E(z)^{-2/3}$ \citep{voi05_rev}. In this work we adopted the characteristic entropy $K_{500}$, as computed by \citet{pratt10}:
\begin{equation}
\begin{split}
    \small K_{500} = 106 ~ \bigg(\frac{M_{500}}{h^{-1}_{70}~10^{14}~\text{M}_{\odot}}\bigg)^{2/3} ~E(z)^{-2/3}~f{_b}^{-2/3}~h_{70}^{-4/3} ~ \text{keV cm}^2,
\end{split}
\label{eq_k500}
\end{equation}
where $f_b$ is the Universal baryon fraction, which we take to be 0.16 \citep{planck_xiii_16}. We will show, through the combined sample of HIGHMz, ESZ, X-COP, and \rexcess\ clusters, that the adopted self-similar scaling may not be adequate to describe the properties of galaxy clusters, due to the impact of non-gravitational processes on the ICM. In Sect. \ref{Am_section} and Appendix \ref{App_Az_dependence}, we will provide updated dependencies on mass and redshift that go beyond self-similar predictions.

Predictions from cosmological simulations that implement gravitational processes only (e.g. \citeauthor{voit+05} \citeyear{voit+05}) can be considered as a baseline to assess the impact of non-gravitational processes on the ICM. These simulations are commonly referred to as `non-radiative' and predict that, in the gravity-dominated regime, cluster entropy  outside the core regions should follow a power law increase out to $\sim 2 ~R_{200}$. Any other physical processes capable of injecting energy into the ICM would result in an overestimate of the ICM entropy with respect to the expected power law behaviour. \citet{voit+05} presented a simple formula that provides both the normalisation and slope of this expected power law, normalised for an overdensity $\Delta = 200$. In this work, we adopt the equivalent equation computed by \citet{pratt10}, written for an overdensity $\Delta = 500$:
\begin{equation}
    K/K_{500} = 1.42 ~(R/R_{500})^{1.1}.
\label{voit_equation}
\end{equation}
Given their notably high masses, one would expect the properties of HIGHMz clusters to be mainly driven by gravitational processes at large radii and so their external entropy profiles to closely follow the prediction by \citet{voit+05}. We will verify if this is the case in the following.

\section{Results}
\label{results}

\subsection{Entropy profiles of HIGHMz clusters}
\label{sec_ent_results}

\begin{figure*}
            \centering
            \hspace{0.245cm}\includegraphics[width=0.78\textwidth]{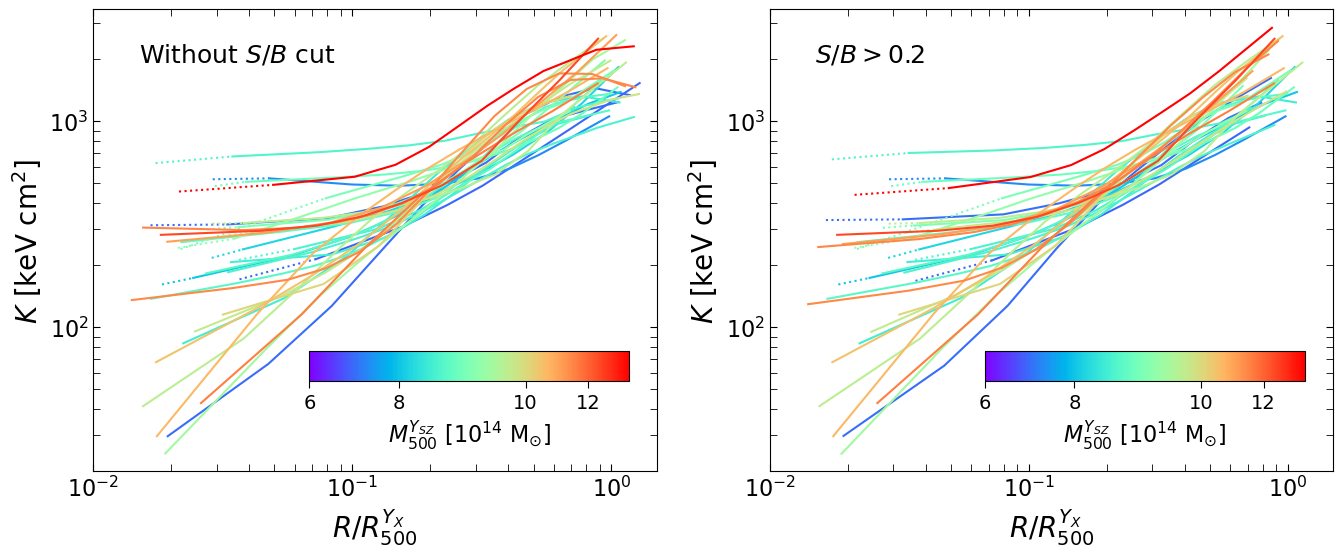}
            \includegraphics[width=0.8\textwidth]{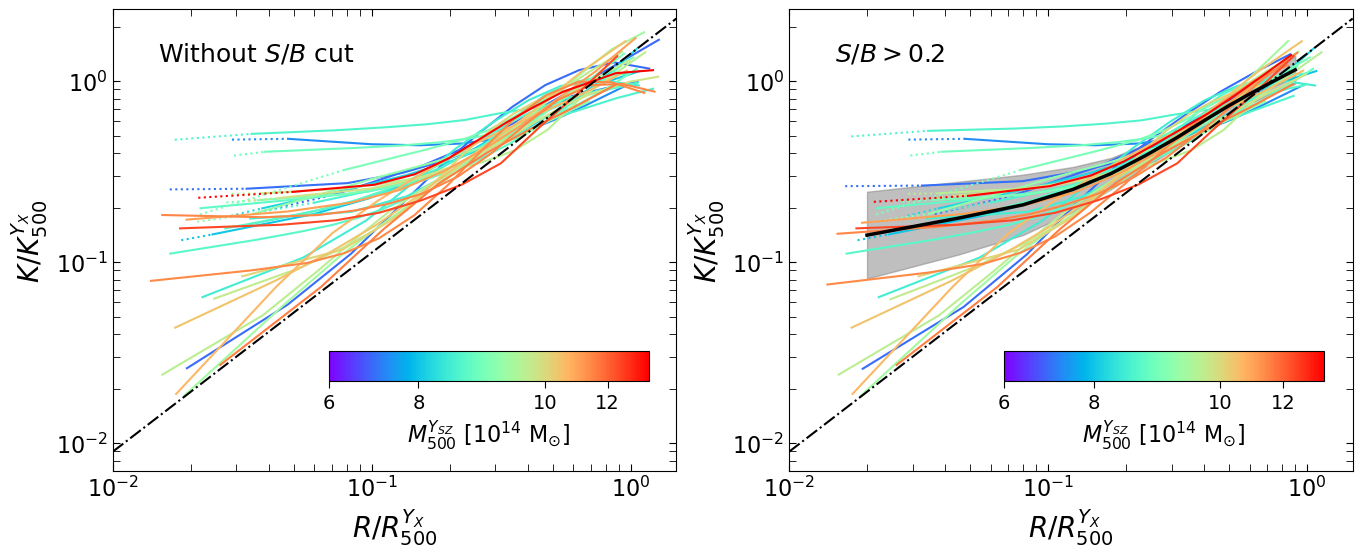}
            \includegraphics[width=0.8\textwidth]{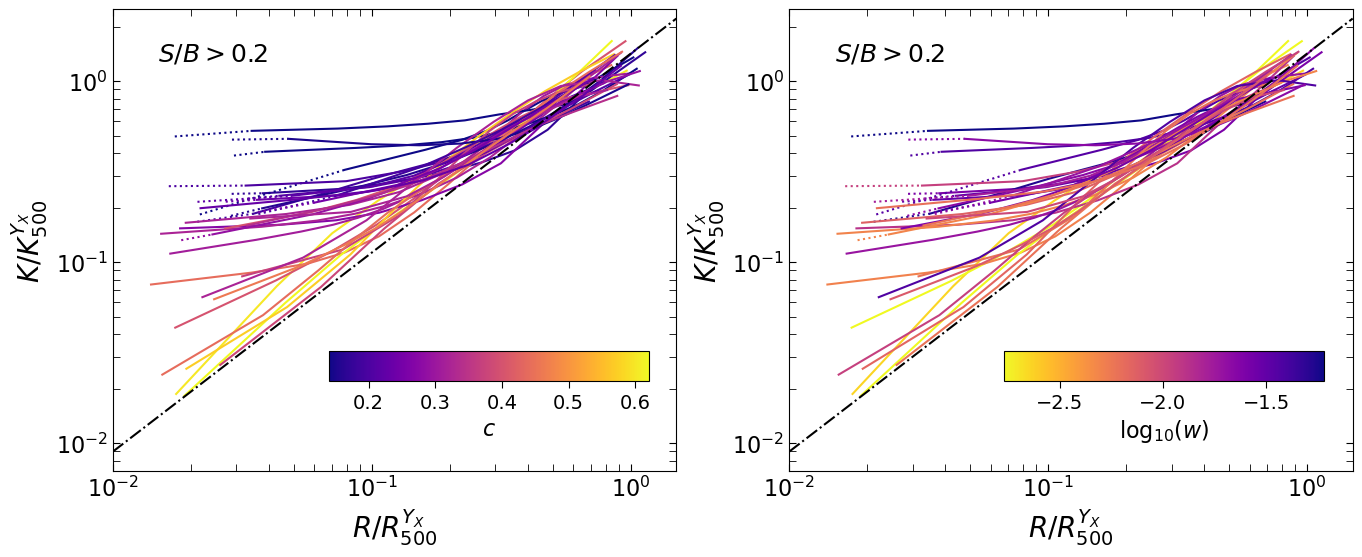}
            \caption{\footnotesize Entropy profiles of HIGHMz clusters, derived using azimuthal median densities. The profiles are shown in physical units in the upper panels, while they are scaled using $K_{500}^{Y_{\rm X}}$ in the centre and bottom. In the upper and central panels, they are colour coded according to the mass, while in the bottom we show the dependence on the concentration (left) and centroid shift (right), as measured by \citet{campitiello22}. The two labels `Without $S/B$ cut' and `$S/B>0.2$' indicate whether all the temperature bins were used in the deprojection or we limited to measurements with $S/B > 0.2$, respectively. In the central right panel, we also plot the median profile (solid black) and the measured intrinsic dispersion (grey shaded area). At small radii, dotted lines denote profiles derived assuming a constant core temperature on the density radial grid. Black dash-dotted lines in the central and lower panels are predictions from non-radiative simulations \citep{voit+05}. }
            \label{K_prof}
\end{figure*}

We present in Fig. \ref{K_prof} the entropy profiles of the HIGHMz clusters, derived from azimuthal median densities. Those obtained from azimuthal mean densities have similar properties and are presented in Appendix \ref{sect_mean_entropy}. To ensure a better examination of the core regions, for those clusters with poor central temperature resolution, we computed the entropy on the density radial grid, assuming a constant central temperature, as already done in previous works (\citeauthor{donahue05} \citeyear{donahue05}; \citeauthor{cavagnolo09} \citeyear{cavagnolo09}; \citeauthor{pratt10} \citeyear{pratt10}). These additional measurements, if any, are shown using dotted lines in Fig. \ref{K_prof}.

In the upper panels of Fig. \ref{K_prof}, the profiles are shown in physical units, with their radii normalised by $R_{500}^{Y_{\rm X}}$. On the left, all the temperature bins have been included in the deprojection, while on the right we have limited the deprojection to temperature bins with $S/B > 0.2$, to reduce the impact of the uncertainty in temperature measurements at large radii discussed in Sect. \ref{T_2D}. As already noted in previous works (e.g. \citeauthor{cavagnolo09} \citeyear{cavagnolo09}; \citeauthor{sun09} \citeyear{sun09}; \citeauthor{pratt10} \citeyear{pratt10}), most profiles do not show simple power-law behaviour down to arbitrarily small radii, but some flattening is typically observed in the central regions $(R \lesssim 0.2~R_{500})$. Here, considerable dispersion is observed, which then reduces moving outwards, where the measured entropy profiles increase with radius following power laws with almost the same slope. At radii $R \gtrsim 0.6~ R_{500}$, the effect of excluding temperature measurements with $S/B < 0.2$ is visible observing the shape of the profiles: some of the clusters exhibit significant flattening in the outskirts when no cut for $S/B$ is applied (top left); conversely, including only temperature bins with $S/B > 0.2$ in the deprojection allows to regularise the shape of the entropy profiles at all radii, especially in the outskirts, where the entropy profiles resume a steadily increasing trend (top right). As the inclusion of low-$S/B$ temperature measurements may introduce a bias into the ICM entropy reconstruction at large radii, as mentioned above, we therefore consider profiles with a cut for $S/B > 0.2$ as our reference. 

In Fig. \ref{K_prof} (upper panels), gas entropy profiles are colour coded according to their values of $M_{500}^{Y_{\rm SZ}}$, with blue to red for least to most massive clusters. This shows that a clear mass dependence is present beyond $R\sim0.2~R_{500}^{Y_{\rm X}}$, with the most massive clusters exhibiting higher entropy, when this is not rescaled for any global quantity. By adopting the self-similar rescaling for $K_{500}^{Y_{\rm X}}$ (Eq. \ref{eq_k500} using $M_{500}^{Y_{\rm X}}$), and thus taking into account the dependence on both cluster mass and redshift, the correlation with cluster mass disappears at large radii (central panels). In the outskirts, we observe perhaps even more clearly the impact of excluding temperature measurements with $S/B < 0.2$ from the analysis on the shape of the profiles. After this correction, the entropy profiles of massive clusters resume a steadily increasing trend with radius also at $\sim R_{500}$ and more closely follow the predictions from pure gravitational collapse (Eq. \ref{voit_equation}, \citeauthor{voit+05} \citeyear{voit+05}), represented in Fig. \ref{K_prof} as the black dash-dotted lines. Conversely, in the core regions, notable scatter is still observed.

In the lower panels of Fig. \ref{K_prof}, the entropy profiles of HIGHMz clusters are colour coded according to their morphological indicators, such as the concentration ($c$, left) and centroid shift ($w$, right), as measured by \citet{campitiello22} and reported in Table \ref{table:1}. In the central regions we find a strong (anti-)correlation with ($c$) $w$, meaning that disturbed clusters tend to have higher central entropy, likely due to residual merger activity, while cool-core clusters exhibit lower central entropy, which is thought to arise from a balance between feedback and cooling processes \citep{pratt10}. Not surprisingly, the dependence of the ICM entropy on the two morphological indicators is similar, although they are sensitive to different cluster scales; indeed, the correlation with $c$, which is directly linked to the core properties, is stronger. Finally, we notice that moving outwards the dependence on morphological parameters is reduced and appears even reversed at $\sim R_{500}$, with cool-core clusters featuring higher, and steeper, entropy profiles, reflecting the behaviour of the gas density at these radii \citep[e.g.][]{maughan12}. A more quantitative estimate of the dependence of the gas entropy on both $c$ and $w$ will be provided in Sect. \ref{sect_corr_c_w}, in comparison with simulated datasets.

We also computed the median and the intrinsic scatter of the profiles, following the procedure detailed in Sect. 6.1 of \citet{bartalucci+23}. As a first step, we interpolated each scaled profile linearly in the log-log plane, on a common grid of twelve bins in the $[0.02 - 1]~R_{500}$ radial range. The medians were then derived through a Monte Carlo technique: for each radial bin, we generated 1000 random realisations of each entropy measurement, normally distributed around its nominal value and adopting the statistical error as sigma. From the final stacked distribution at each radius, we then computed the median and associated errors as the $50^{\text{th}}$ and $[16^{\text{th}}-84^{\text{th}}]$ percentiles, respectively. To derive the intrinsic scatter of the entropy measurements as a function of radius, we fitted the interpolated data points at each radial bin in a Bayesian framework, using PyMC v.5.6.1 \citep{abril-pla23}, with the model:
\begin{equation}
\label{eq1}
    K/K_{500}^{Y_{\rm X}} = A\cdot \text{exp}(\pm \sigma_{\mathrm{int}}),
\end{equation}
where $A$ is a constant and $\sigma_{\mathrm{int}}$ the intrinsic scatter. Note that the best-fitting values of $A$ and $\sigma_{\mathrm{int}}$ are different at each considered bin. We report our results in Table \ref{tab:median} and plot the median profile both in the central right panel of Fig. \ref{K_prof} and also in Fig. \ref{scalings}, where we compare it with other scalings adopted throughout the paper.
\begin{table}
    \centering
    \caption{\footnotesize Median entropy profile and intrinsic dispersion of HIGHMz clusters.}
    \begin{tabular}{ccc}
    \toprule
    \toprule
        $R/R_{500}^{Y_{\rm X}}$ & $K/K_{500}^{Y_{\rm X}}$ & $\sigma_{\mathrm{int}}$\\
    \midrule
        $0.02$ &  $0.14 \pm 0.01$ & $0.55 \pm 0.10$\\
        $0.05$ &  $0.18 \pm 0.01$ & $0.47 \pm 0.09$\\
        $0.08$ &  $0.21 \pm 0.01$ & $0.38 \pm 0.07$\\
        $0.13$ &  $0.25 \pm 0.01$ & $0.29 \pm 0.05$\\
        $0.18$ &  $0.31 \pm 0.01$ & $0.20 \pm 0.03$\\
        $0.24$ &  $0.39 \pm 0.01$ & $0.14 \pm 0.02$\\
        $0.31$ &  $0.49 \pm 0.01$ & $0.13 \pm 0.02$\\
        $0.40$ &  $0.60 \pm 0.01$ & $0.12 \pm 0.02$\\
        $0.50$ &  $0.73 \pm 0.01$ & $0.12 \pm 0.02$\\
        $0.60$ &  $0.85 \pm 0.02$ & $0.12 \pm 0.02$\\
        $0.70$ &  $0.98 \pm 0.02$ & $0.14 \pm 0.02$\\
        $0.90$ &  $1.15 \pm 0.03$ & $0.18 \pm 0.03$\\
    \bottomrule
    \end{tabular}
    \tablefoot{The median and the dispersion are measured using Eq. \ref{eq1} and are shown in Fig. \ref{K_prof} (central right panel). The median profile is also plotted in Fig. \ref{scalings}, as black dots.}
    \label{tab:median}
\end{table}

\subsection{Analytical fits}
\label{codes_median}
\begin{figure}
\centering
            \includegraphics[width=0.43\textwidth]{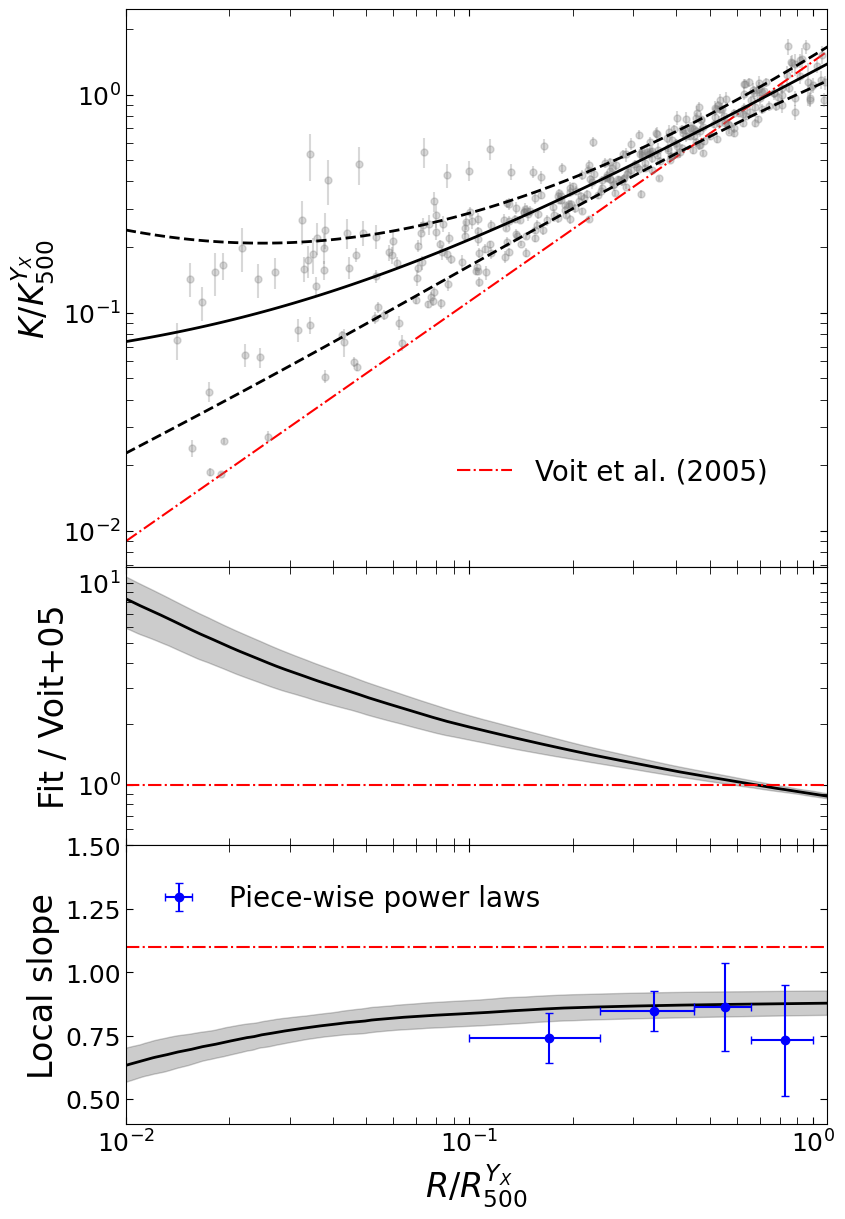}
            \caption{\footnotesize Joint fit of the HIGHMz entropy profiles using the constant plus power-law parameterisation (Eqs. \ref{eq_an_fit} and \ref{eq_sct_an_fit}). Top panel: best-fitting model (black), super-imposed to observational measurements (grey dots). Black dashed lines mark the intrinsic scatter, while red dash-dotted line is the prediction from non-radiative simulations \citep{voit+05}. Central panel: ratio between our best-fitting model and predictions from non-radiative simulations. Grey shaded area is the associated statistical error. Bottom panel: local slope of the best-fitting analytical model (black), together with the associated statistical error (grey). In blue are the slopes obtained using fitting with piece-wise power laws (Eq. \ref{eq_piecewise_pl}). Red dash-dotted line is the canonical $1.1$ slope.}
            \label{analytical_fit}
\end{figure}

\begin{table}
\caption{\footnotesize Best-fitting parameters of the functional forms describing HIGHMz entropy profiles.}
\centering                          
\begin{tabular}{ccccc}  
\multicolumn{5}{c}{\small Global analytical fit \normalsize}\\
\toprule   
\toprule
\multicolumn{2}{c}{Parameter}& \multicolumn{2}{c}{Prior} &Best fit \\
\midrule  
\multicolumn{2}{c}{log($K_0$)} & \multicolumn{2}{c}{$U(-7,0)$} &\hspace{-0.21cm}$-2.97 \pm 0.37$\\
\multicolumn{2}{c}{$K_1$} & \multicolumn{2}{c}{$U(1,2)$} & $1.23 \pm 0.03$ \\
\multicolumn{2}{c}{$\alpha$} & \multicolumn{2}{c}{$U(0,2)$}  & $0.87 \pm 0.04$\\
\midrule
\multicolumn{2}{c}{$\sigma_0$} & \multicolumn{2}{c}{$U(0,0.5)$} & $0.12 \pm 0.01$\\
\multicolumn{2}{c}{$\sigma_1$} & \multicolumn{2}{c}{$U(0,0.5)$} & $0.07 \pm 0.02$\\
\multicolumn{2}{c}{$x_0$} & \multicolumn{2}{c}{$U(0,1)$} & $0.44 \pm 0.05$\\
\midrule      
&&&& \\
\multicolumn{5}{c}{\small Piece-wise power-law fits \normalsize} \\
\midrule   
\midrule
$x_{\mathrm{in}}$ & $x_{\mathrm{out}}$ & $A$ & $B$ & $\sigma_{\mathrm{int}}$ \\
&&$U(0,2)$ & $U(0,2)$ & $HC(1)$\\
\midrule
$0.10$ & $0.24$ & $1.14 \pm 0.18$& $0.74 \pm 0.10$& $0.22 \pm 0.02$\\
$0.24$ & $0.45$ & $1.32 \pm 0.09$& $0.85 \pm 0.08$& $0.11 \pm 0.01$\\
$0.45$ & $0.66$ & $1.33 \pm 0.11$& $0.86 \pm 0.17$& $0.13 \pm 0.02$\\
$0.66$ & $1.00$ & $1.25 \pm 0.06$& $0.73 \pm 0.22$& $0.15 \pm 0.02$\\
\bottomrule
\end{tabular}
\tablefoot{Parameters in the upper Table refers to the constant plus power-law model (Eqs. \ref{eq_an_fit} and \ref{eq_sct_an_fit}), while those in the lower Table to the piece-wise power laws (Eq. \ref{eq_piecewise_pl}). We assumed flat priors for all the parameters, except for $\sigma_{\mathrm{int}}$, for which we took a Half-Cauchy distribution with $\beta=1$.}  
\label{table:fit}
\end{table}
To provide further information on the properties of HIGHMz entropy profiles, including their normalisation, slope and intrinsic scatter at all radii, we adopted a constant plus power-law model, as first introduced by \citet{donahue05}, together with a radially dependent scatter (Eq. 5 of \citeauthor{ghirardini19} \citeyear{ghirardini19}). This parameterisation is able to capture both the power-law increase of the profiles at large radii and the flattening observed in the central regions. By using PyMC, we jointly fitted the scaled entropy profiles in the radial range $[0.01-1]~R_{500}^{Y_{\rm X}}$, with the model:
\begin{equation}
    K(x)/K_{500}^{Y_{\rm X}} = (K_0 + K_1 ~x^{\alpha}) \cdot \text{exp}[\pm \sigma_{\mathrm{int}}(x)],
    \label{eq_an_fit}
\end{equation}
where $x = R/R_{500}^{Y_{\rm X}}$ and $K_0$ is the central entropy excess above the power law, with normalisation $K_1$ and slope $\alpha$, that is measured at large radii; $\sigma_{\mathrm{int}}(x)$ is the intrinsic scatter\footnote{We emphasise that what we call
‘intrinsic scatter’ is an upper limit to the true dispersion
of the profiles, as hidden systematic uncertainties may lead to an overestimate of the real scatter.} of the profiles and varies with the radius following the quadratic functional form:
\begin{equation}
    \sigma_{\mathrm{int}}(x) = \sigma_0 + \sigma_1 ~ \text{log}^2(x/x_0),
    \label{eq_sct_an_fit}
\end{equation}
with $\sigma_1$ the width of the log-parabola, and $x_0$ and $\sigma_0$ the location and the intercept of the minimum of the log-parabola, respectively. The adopted priors and the resulting best-fitting parameters are reported in Table \ref{table:fit} (top); in Fig. \ref{analytical_fit}, we show our best-fitting curve (top), with particular focus on its normalisation (centre) and slope (bottom), in comparison with predictions from \citet{voit+05}. 

Within $\sim 0.1 ~R_{500}^{Y_{\rm X}}$ the best-fitting analytical profile becomes progressively flatter and the measured dispersion is large, as already discussed; moving out to the cluster outskirts, the intrinsic scatter is reduced and the best-fitting profile approaches the predictions from pure gravitational collapse \citep{voit+05}, albeit with a gentler slope at all radii $(\alpha \sim 0. 9$ at $R_{500}^{Y_{\rm X}})$. The measured best-fitting profile is above the theoretical predictions for $R \lesssim 0.7~R_{500}^{Y_{\rm X}}$, while it becomes lower at larger radii. In particular, the measured gas entropy is $\sim 90\%$ of that predicted by \citet{voit+05} at $R_{500}^{Y_{\rm X}}$.
\begin{figure*}
            \centering
            \includegraphics[width=0.41\textwidth]{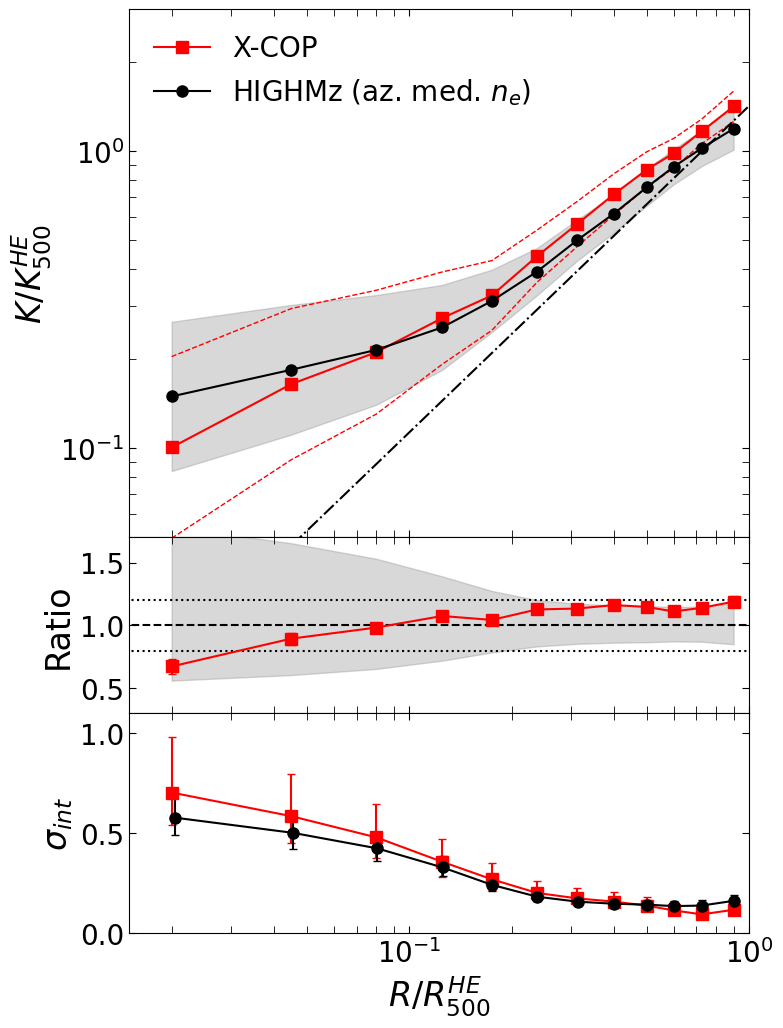}\hspace{0.5cm}
            \includegraphics[width=0.41\textwidth]{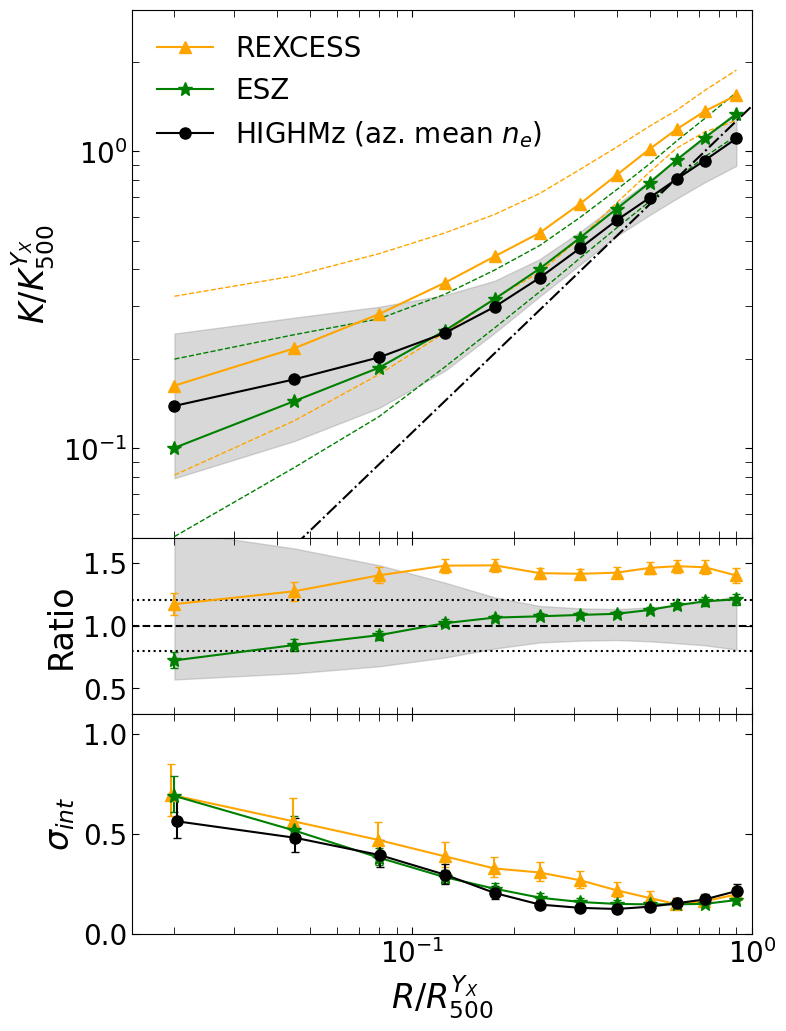}
            \caption{\footnotesize Comparison with other observational samples. On the left, we compare our results to X-COP (red squares), while on the right to \rexcess\ (yellow triangles) and ESZ (green stars) clusters. Top: median entropy profiles, with their intrinsic scatter. Centre: ratio to the HIGHMz median profile. The grey shaded area is the intrinsic scatter of the HIGHMz clusters, while the horizontal dotted lines indicate a $20\%$ variation. Bottom: radial profiles of intrinsic scatter.}
            \label{med_samples}
\end{figure*}

Due to the rigidity of the adopted analytical model, any possible variations of the best-fitting slope with the radius would be smoothed away in favour of a mean slope. To address this issue, we also tested the fit with piece-wise power laws (as described in Eq. 4 of \citeauthor{ghirardini19} \citeyear{ghirardini19}), which allow for greater flexibility. We divided the profiles into four radial bins, in the range $[0.1-1]~R_{500}^{Y_{\rm X}}$, and used PyMC again to build the model:
\begin{equation}
    K(x)/K_{500}^{Y_{\rm X}} = A\cdot x^B \cdot \text{exp}(\pm \sigma_{\mathrm{int}}),
    \label{eq_piecewise_pl}
\end{equation}
where $A$ is the normalisation of the power law, with slope $B$, and $\sigma_{\mathrm{int}}$ is the intrinsic scatter in each bin. The priors and best-fitting parameters are reported in Table \ref{table:fit} (bottom), while the measured slopes are also shown in Fig. \ref{analytical_fit} (bottom). We do not find significant variations of the slopes with radius, which are in agreement with results from the global analytical fit. The measured piece-wise power-law slopes are flatter than the canonical 1.1 value at all radii.


\subsection{Comparison with other observational samples}
\label{other_samples}

We compared HIGHMz entropy profiles to those of other observational samples found in the literature, namely \rexcess, ESZ, and X-COP (see Sect. \ref{sample_presentation}). We stress that a blind comparison of the products obtained from different samples, analysed with distinct techniques, is likely to be affected by the distinct systematic uncertainties of each sample results. In the following, we take this into account and perform two separate comparisons, bringing the four samples to conditions that are as similar as possible:
\begin{itemize}
    \item the entropy profiles of X-COP clusters were derived using azimuthal median densities and rescaled (for $K_{500}$ and $R_{500}$) using hydrostatic equilibrium masses \citep{eckert22}. Consequently, we compare these data with HIGHMz entropies derived and scaled in a similar way;
    \item similarly, \rexcess\ and ESZ profiles were produced using azimuthal mean densities and rescaled using $M_{500}^{Y_{\rm X}}$ \citep{pratt10}. For comparison, we thus consider HIGHMz profiles derived from mean densities and adopt masses derived from the $Y_X$ proxy in the rescaling. These profiles are also shown in Fig. \ref{K_prof_mean}. 
\end{itemize}
Also for X-COP, \rexcess, and ESZ, we adopt a rescaling by $K_{500}$ following Eq. \ref{eq_k500}, using $f_b = 0.16$, as already done for HIGHMz clusters. For each of the four samples, we computed the median entropy profile and the intrinsic scatter following the method outlined in Sect. \ref{sec_ent_results}. The results are illustrated in Fig. \ref{med_samples}, where we compare HIGHMz to X-COP (left) and to \rexcess\ and ESZ (right). The two HIGHMz median profiles shown in Fig. \ref{med_samples} are also compared in Fig. \ref{scalings} to the `reference' measurement, i.e. entropy using median densities and $M_{500}^{Y_{\rm X}}$ in the rescaling.

The median scaled entropy profile of HIGHMz consistently lies below those of the other samples, regardless of whether entropies are derived from azimuthal median (left) or mean (right) densities, and regardless the adopted mass in the scaling. The only exception is given by the core region, where we measure flatter entropies. This may be due to a combination of different effects, such as a different morphological distribution of HIGHMz clusters and/or a distinct resolution of the profiles due to the different redshift range. Further discussion on the distribution of the central entropies is provided in Sect. \ref{central_distr}. 

Beyond $\sim 0.1~R_{500}$, the median profiles of the four samples exhibit similar shapes, as shown by the nearly constant ratios (sample/HIGHMz) with radius in the central panels of Fig. \ref{med_samples}. The good agreement in shape with X-COP suggests that restricting the analysis to temperature measurements with $S/B>0.2$ for HIGHMz has a comparable impact on the entropy profiles as a joint X/SZ analysis, as also pointed out by \citet{rossetti24} for the temperatures. Conversely, larger discrepancies are found when comparing normalisations. \rexcess, X-COP, and ESZ median profiles are, on average, $\sim42\%$, $\sim14\%$ and $\sim12\%$ higher than those of HIGHMz, respectively. These differences in normalisation correlate with the median masses of the samples (Fig. \ref{samples}): low mass objects feature higher scaled entropy profiles than massive systems, which more closely approach the predictions from non-radiative simulations, consistent with the idea that non-gravitational effects play a more important role at the low mass end. We discuss further the departures from self-similar predictions in Sect. \ref{Am_section}. 

In Fig. \ref{med_samples} (bottom), we also show the dispersion of the entropy profiles for the four samples as a function of the radius. In general, the measured intrinsic scatter profiles are in agreement at all radii. We measure considerable dispersion in the central regions, reflecting the morphological variety of the clusters, while the scatter reduces towards the external regions, where clusters appear to be more self similar. A small increase of the intrinsic scatter is then observed at radii $R \gtrsim 0.8~R_{500}$, especially using the azimuthal mean, likely due to the larger impact of undetected sub-clumps. The only notable exception regards \rexcess\ clusters, which display a larger scatter (factor of $\sim 1.5$ with respect to ESZ and HIGHMz) in the radial range $[0.1-0.4]~R_{500}$, due to the wide range of masses that are present in the sample. To test this hypothesis, we divided \rexcess\ into two equal parts based on cluster masses and verified that the measured scatter of the more massive half is consistent with the ESZ and HIGHMz values.

\subsection{Comparison with simulations}
\label{simulations}

We now compare the HIGHMz entropy profiles with those of clusters taken from the MACSIS and The300 datasets (see Sect. \ref{sim_presentation}), to test whether our theoretical understanding of the physical processes acting at the galaxy cluster scale can reproduce the properties of the observed entropy profiles. 

As specified in Sect. \ref{sim_presentation}, we assumed that $Planck$ masses present an average $20\%$ hydrostatic mass bias. We therefore consider HIGHMz masses corrected accordingly, i.e. $M_{500}^{Y_{\rm SZ}}/(1-b)$, with $1-b=0.8$, to compute both $K_{500}$ and $R_{500}$, which we then used in the rescaling. To ensure a fair comparison, we also adopted the scaling for $K_{500}$ reported in Eq. (\ref{eq_k500}) for the simulations, using $f_b = 0.16$, as done for observations. We note that we have used mass-weighted entropy profiles for the simulated clusters, to reduce the effect of cold and dense sub-structures that are not masked in the simulations. However, in Appendix \ref{App_sim} we also present spectroscopic-like \citep{mazzotta04} entropy profiles, as compared to the mass-weighted ones (Fig. \ref{Tmw_sl_sim}).

\subsubsection{Shape and dispersion}
\label{sect_shape_sim}
\begin{figure}
\centering
            \includegraphics[width=0.42\textwidth]{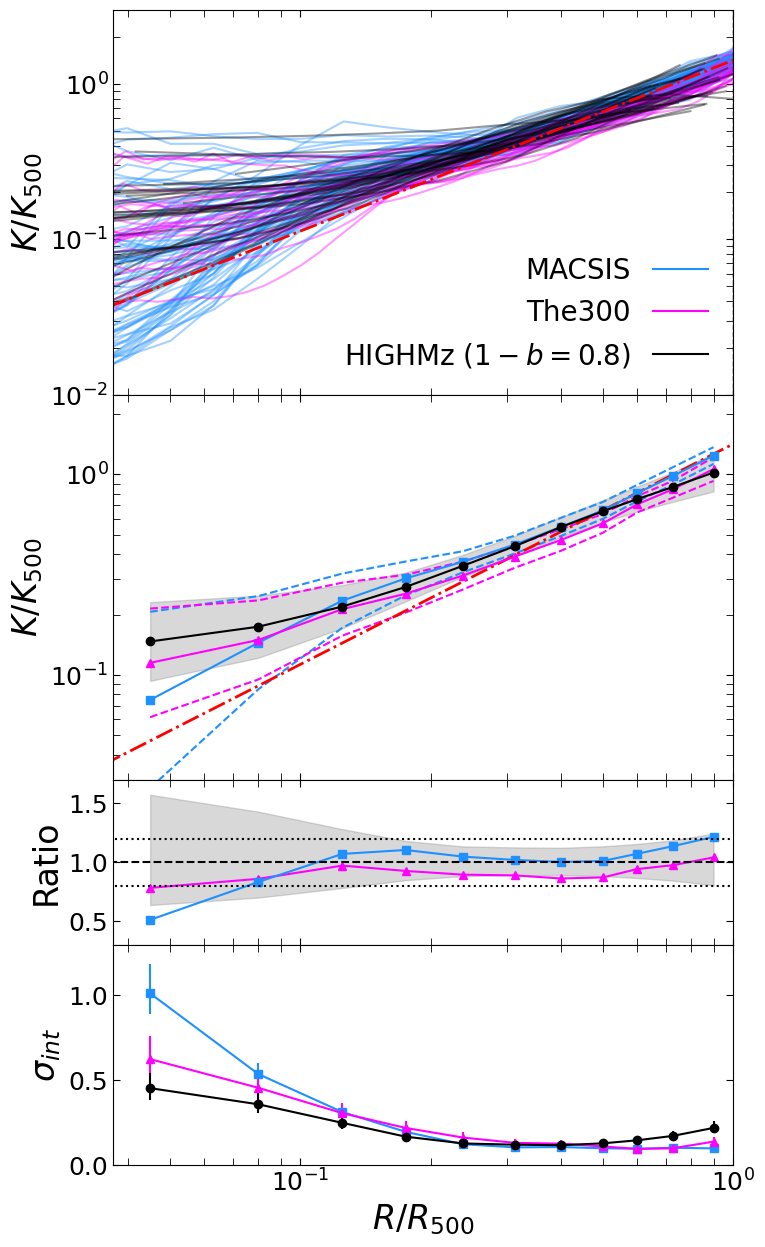}
            \caption{\footnotesize Comparison with simulated datasets. From the top to the bottom: i) measured profiles for MACSIS (light blue), The300 (violet), and HIGHMz, rescaled to mimic a $20\%$ hydrostatic bias, as described in the text (black); ii) median entropy profiles of the three samples, together with their intrinsic scatters and the red dash-dotted line showing the prediction from \citet{voit+05}; iii) ratio to HIGHMz, with the grey shaded area showing the intrinsic dispersion of HIGHMz entropies and horizontal dotted lines marking a $20\%$ discrepancy; iv) radial profile of the intrinsic dispersion for the three samples.}
            \label{K_sim}
\end{figure}

Figure \ref{K_sim} compares the scaled entropy profiles of the HIGHMz, MACSIS, and The300 samples. In general, the simulations reproduce the observed dispersion in the central regions and also the steady increase of ICM entropy with radius, with reduced scatter. Beyond the core, both MACSIS and The300 profiles increase according to a power law, with almost the same slope predicted from pure gravitational collapse \citep{voit+05}. For a more quantitative comparison, we measured the median and the intrinsic scatter of the entropy profiles as a function of radius, following the same procedure detailed in Sect. \ref{sec_ent_results}.
\begin{figure*}
\centering
            \includegraphics[width=0.92\textwidth]{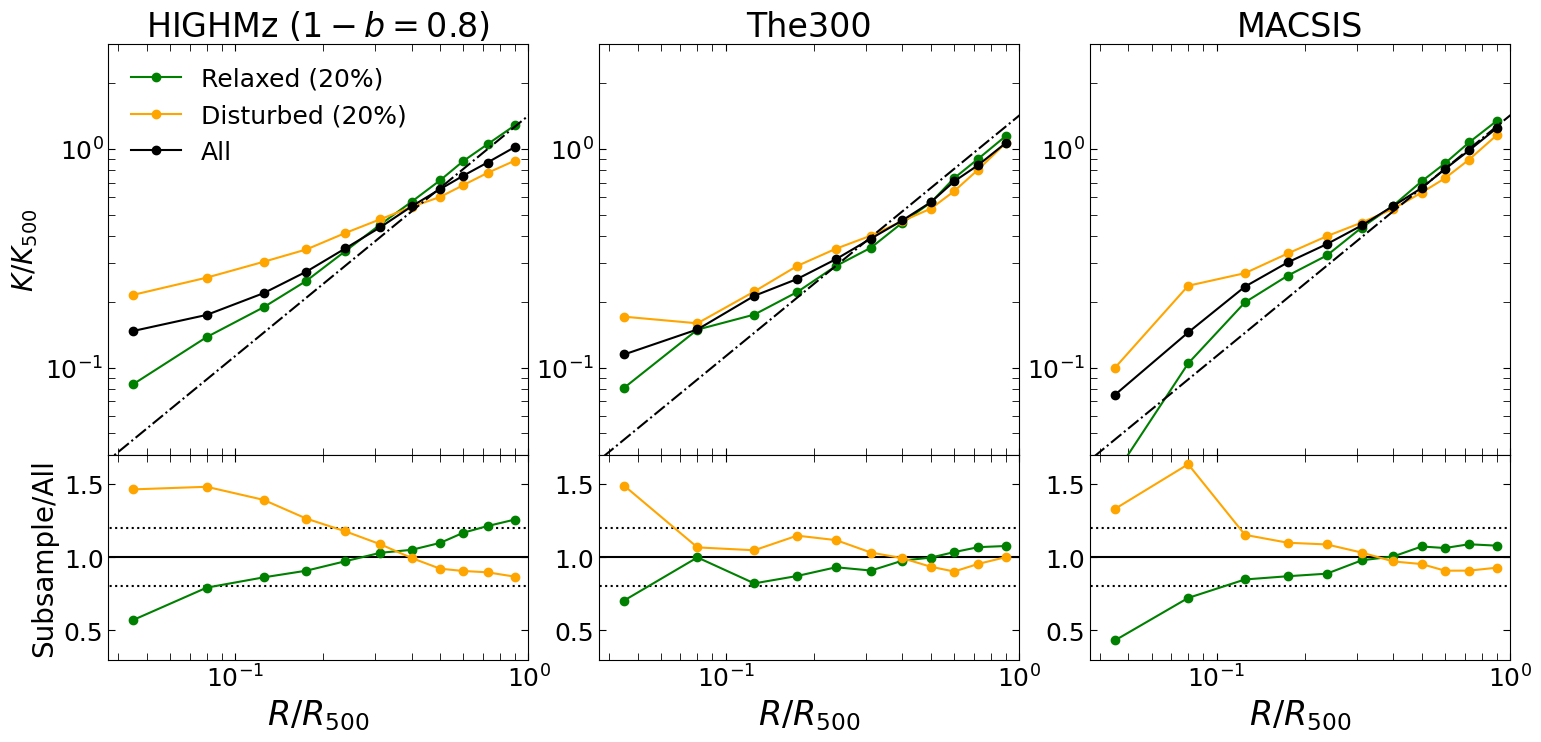}
            \caption{\footnotesize Comparison between relaxed and disturbed systems, for HIGHMz (left), The300 (centre), and MACSIS (right). Green is used for the most relaxed clusters, while yellow for most disturbed ones. Black is used for the median profile obtained using all the clusters in each sample. In the bottom panels, we show the ratios to the medians of the full samples, with the dotted lines marking a $20\%$ discrepancy.}
            \label{subsample_sim}
\end{figure*}

Beyond the core, we find a good agreement in shape with the median entropy profile of both simulated datasets, although HIGHMz profiles are flatter in the outskirts. Regarding the normalisation, HIGHMz median profile lies between those of MACSIS and The300, which slightly overestimate ($\sim 6\%$) and underestimate ($\sim 9\%$) the observed one, while still remaining within the measured dispersion at all radii. These differences are due to the combination of differences in temperatures and electron densities, as shown in Appendix \ref{App_sim}. Looking at the central regions ($R \lesssim 0.1 ~R_{500}$), larger discrepancies are observed for MACSIS clusters, which predict strong cooling in the core; as a result, the simulated galaxy clusters have low central temperatures and peaked central densities (see Fig. \ref{T_ne_sim}), which together give a systematically lower gas entropy than observed. This is not a new finding, as a similar behaviour has already been observed in previous works comparing simulations with observations, \citep[e.g.][]{barnes17} and also a recent study of FLAMINGO clusters \citep{braspenning24}, based on a similar code. Conversely, somewhat  better agreement with observations is observed in the core of The300 clusters, probably favoured by the implemented artificial thermal diffusion term \citep{rasia15}, which allows for more efficient gas mixing in the central regions.

Regarding the dispersion of the profiles, presented in Fig. \ref{K_sim} (bottom), the simulations are able to reproduce the general behaviour observed in real clusters, albeit with some differences. For example, both simulated datasets overestimate the observed intrinsic dispersion in the central regions, although the difference is not statistically significant for The300, and underestimate it in the outskirts. At intermediate radii, i.e. $[0.2-0.5]~R_{500}$, both simulated datasets reproduce well the measured dispersion for HIGHMz clusters.

\subsubsection{Correlation with morphological parameters}
\label{sect_corr_c_w}

In Fig. \ref{K_prof}, we have highlighted that the central dispersion of HIGHMz entropy profiles is closely related to their dynamical state, as supported by the strong correlation between entropy and morphological indicators, such as $c$ and $w$, and that this dependence is then reduced outside the core. In the following, we test the ability of simulations to reproduce the dependence of the entropy profiles on $c$ and $w$. We approach this question in two different ways: by dividing the samples into disturbed and relaxed objects and comparing their median profiles; and by quantifying the correlation between entropy and morphological indicators as a function of radius.

We first divided HIGHMz, MACSIS, and The300 clusters into two sub-samples based on their classification as either most relaxed or most disturbed. This was achieved by sorting the clusters according to their centroid shift value, $w$, and selecting those with $w$ values falling in the lower and upper tails of the distribution. Specifically, we identified the most relaxed and disturbed clusters as the $20\%$ of the clusters with the lowest and highest $w$, respectively. This approach only considers clusters that are in the two tails of the distribution, thus reducing the impact of potential systematics arising from differences in the calculation of $w$ between observations and simulations. We then computed the median entropy profiles of the two sub-samples and compared them with the median profile of all the clusters in HIGHMz, MACSIS, and The300 (Fig. \ref{subsample_sim}).  We find that the simulations reproduce the general behaviour seen in observations, i.e. relaxed clusters have steeper profiles at all radii than perturbed clusters. The difference between the median profiles of the two sub-samples is particularly pronounced in the central regions, where disturbed systems show higher entropy profiles than relaxed ones; the dichotomy disappears at $\sim0.3-0.4~ R_{500}$ and is then slightly reversed in the outer regions, where we observe relaxed clusters more closely approaching predictions from \citet{voit+05}. In the outskirts, the simulations underestimate the differences between the two sub-samples, which is not surprising given the lower reproduced scatter in these regions with respect to the observations, as already noted in Sect. \ref{sect_shape_sim}.
\begin{figure*}
\centering
            \includegraphics[width=0.89\textwidth]{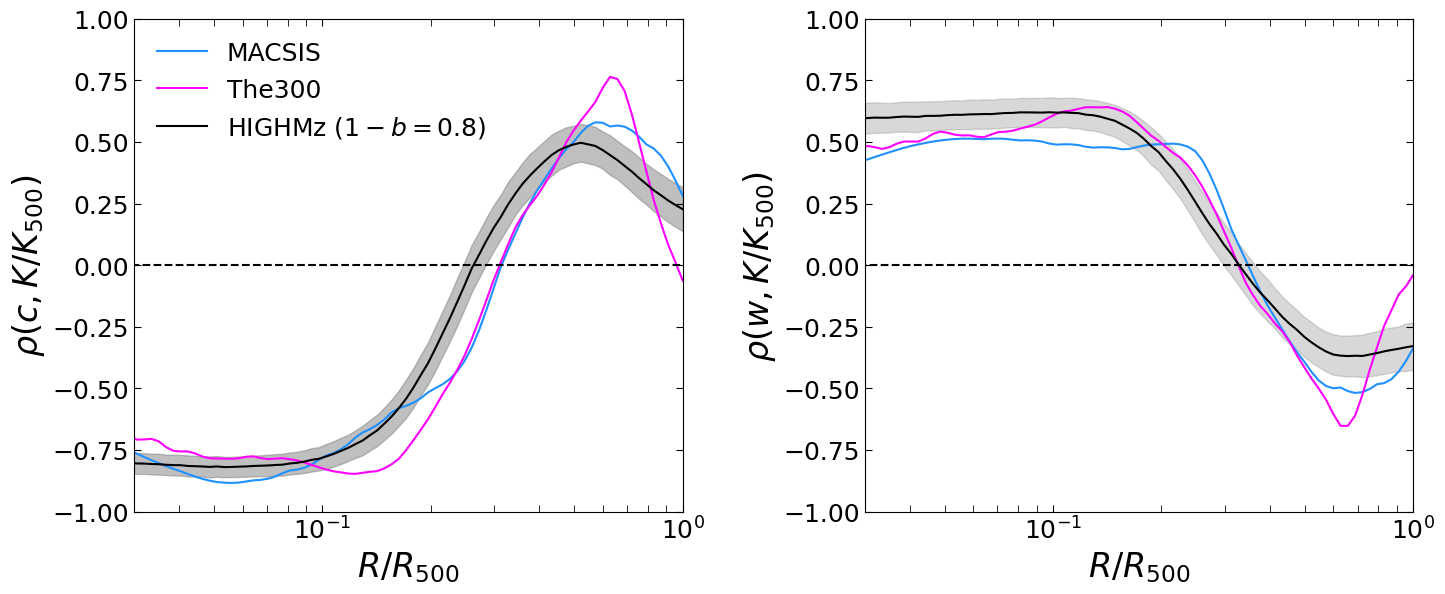}
            \caption{\footnotesize Pearson's correlation coefficient between $c$ (left) or $w$ (right) and scaled entropy $K/K_{500}$ as a function of cluster radius. Light blue, violet and black are used for MACSIS, The300, and HIGHMz, respectively. The grey shaded area is the statistical error associated to HIGHMz measurements.}
            \label{perason_corr}
\end{figure*}

We also calculated a more quantitative comparison of the dependence of the scaled entropy on the morphological parameters for the three samples. Specifically, we interpolated the entropy profiles on a dense radial grid, and for each radius we calculated (in the log-log plane) the correlation between the ICM entropy and both $c$ and $w$, as quantified by the Pearson's $\rho$ parameter. For HIGHMz, we included information on the statistical errors in the measurements by performing a Monte Carlo simulation: for each radius, we generated 1000 random values (around the nominal measurements, with the sigma given by the statistical errors) for $K/K_{500}$, $c$ and $w$, and for each realisation we calculated the correlation. Finally, from these distributions of measured Pearson's $\rho$, we computed the medians and associated errors. Plotting the results as a function of radius as in Fig. \ref{perason_corr} (left for $c$ and right for $w$), it is possible to investigate how the dependence varies from the centre to the outskirts.  Figure~\ref{perason_corr} reflects some of the aspects already discussed in Sect. \ref{sec_ent_results}: in the centre we observe a strong (anti-)correlation with ($c$) $w$ for HIGHMz clusters, with $\rho \sim (-0.8)$ $0.6$. The dependence on $c$ and $w$ reaches a minimum between $R \sim 0.2-0.3~R_{500}$ and finally undergoes a slight inversion at larger  radii. The simulations reproduce both the shape and intensity of these correlations at all radii, indicating that they can accurately replicate the spread of the entropy profiles and the dependence on dynamical state. The only difference is found at $\sim R_{500}$ for The300 where no significant correlation is found. This is consistent with the finding of Fig. \ref{subsample_sim} where the profiles of the relaxed and disturbed sub-samples of these 25 clusters from The300 are almost coincident at $R_{500}$.

\section{Discussion}
\label{discussion}

\subsection{Flattening of the profiles in the cluster outskirts?}
\label{sect_flattening}

The shape of the entropy profiles in the cluster outskirts has been the subject of debate during the past decades, as conflicting results have been derived. According to cosmological simulations which implement gravitational processes only, entropy should exhibit a steady increase with radius ($\propto R^{1.1}$) at least out to $\sim 2~R_{200}$ (\citeauthor{tozzi01} \citeyear{tozzi01}; \citeauthor{voit+05} \citeyear{voit+05}). However, observations exploiting the low particle background of \textit{Suzaku} have indicated significant flattening at $\sim 0.6~R_{200}$ and even at inner radii (e.g. \citeauthor{walker12} \citeyear{walker12}; \citeauthor{urban14} \citeyear{urban14}; \citeauthor{simionescu17} \citeyear{simionescu17}). Various explanations have been discussed to account for the observed flattening in \textit{Suzaku} data, including differences between electron and ion temperatures in the outer ICM regions (\citeauthor{hoscino10} \citeyear{hoscino10}; \citeauthor{akamatsu11} \citeyear{akamatsu11}) and a weakening of the accretion shock as it expands (\citeauthor{lapi10} \citeyear{lapi10}; \citeauthor{cavaliere11} \citeyear{cavaliere11}; \citeauthor{walker12} \citeyear{walker12}). Constraining the radius at which the flattening starts, and the extent of it, would represent a significant step forwards in our understanding of the physical mechanisms occurring at the accretion shock.

New insights were provided by \citet{ghirardini19}, who reconstructed the entropy profiles of X-COP clusters, revealing a steady increase with radius out to at least $\sim R_{200}$, consistent with predictions from simulations. X-COP introduced several methodological improvements, such as the use of median instead of mean densities. In particular, the ability of median densities to exclude clumps down to scales of $\sim 10-20$ kpc in the outer regions has been suggested as the main reason for discrepancies with \textit{Suzaku}. The mean densities, combined with the low resolution of \textit{Suzaku} ($\sim 2$ arcmin), would have prevented the exclusion of cool, over-dense structures, leading to a bias towards higher gas density and lower temperature values, and consequently resulting in an underestimation of the measured gas entropy.

With the high data quality of the CHEX-MATE clusters, we are now able to provide further conclusions about the shape of the entropy profiles in the cluster outskirts. We showed that flattening is occasionally observed at $\sim R_{500}$, both using azimuthal median (Fig. \ref{K_prof}) and mean (Fig. \ref{K_prof_mean}) densities, when all the temperature bins are used in the reconstruction of the gas entropy. Conversely, correcting for the bias highlighted by \citet{rossetti24}, i.e. excluding temperature measurements with $S/B < 0.2$, the entropy profiles resumed a steadily increasing trend, consistent with X-COP and with predictions from cosmological simulations. This suggests that the contribution of azimuthal median densities alone may not fully explain the differences with \textit{Suzaku}, as it was suggested by \citet{ghirardini19}, but a correction of a potential bias in the temperature measurement is also needed.

\begin{figure}
\centering
            \includegraphics[width=0.45\textwidth]{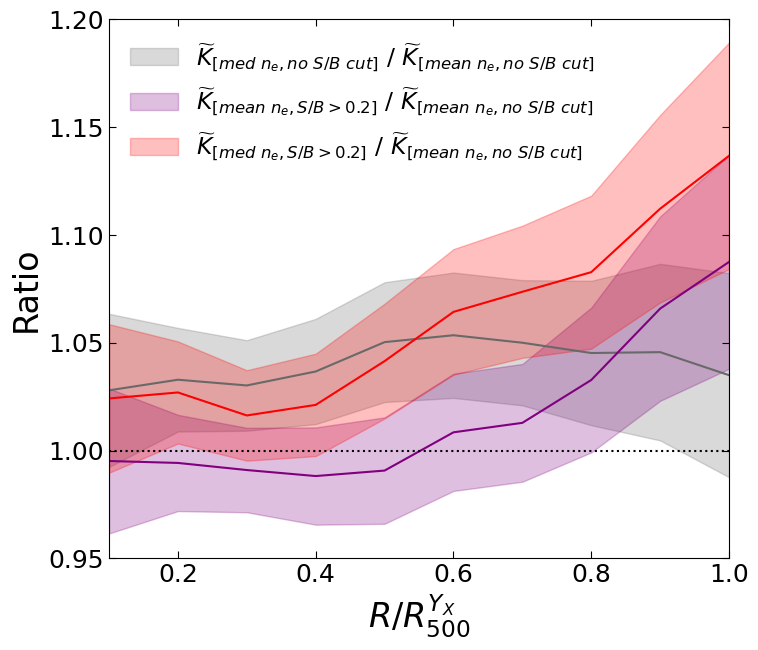}
            \caption{\footnotesize Relative contribution of potential density and temperature-related systematic uncertainties to the median entropy profile. We show the ratios between the medians $\widetilde{K}$ of the scaled entropy profiles, reconstructed using: i) median and mean densities, without cut for $S/B > 0.2$ (grey); ii) mean densities, with and without cut for $S/B > 0.2$ (purple) and iii) a combination of the two (red), as detailed in the text. Shaded areas are the $1\sigma$ statistical errors.}
            \label{ratio_flat}
\end{figure}

We investigate the impact of alleviating the bias in gas density (using the azimuthal median) and temperature (imposing $S/B > 0.2$) separately in Fig. \ref{ratio_flat}. More specifically, we performed the ratio between the median entropy profiles of the sample reconstructed in two different ways:
\begin{itemize}
    \item using azimuthal median and mean densities, with no correction for the bias in temperature (grey). This highlights the possible impact of a more efficient exclusion of cold and over-dense regions from the reconstruction of the gas density, while temperatures are left untouched;
    \item using azimuthal mean density profiles, combined with temperature profiles with and without low-$S/B$ measurements (purple). This shows the impact of correcting the bias in temperature only.
\end{itemize}
In Fig. \ref{ratio_flat}, we show the ratios of the medians of the measured entropy profiles, together with their statistical errors. 

The use of azimuthal median instead of mean density profiles results in a marginal increase ($\lesssim 5\%$) of the ICM entropy at all radii, with a measured trend similar to the one shown in \citet{eckert15} and \citet{bartalucci+23}. Conversely, excluding temperature bins where $S/B<0.2$ barely affects the reconstructed entropy profiles within $\sim 0.6 ~R_{500}$ (although some differences are noted, due to deprojection effects), while it has larger impact in the outer regions ($\sim 8 \%$ at $R_{500}$), especially for those clusters with significantly biased low external temperature measurements. This clearly indicates that, in the outskirts, correcting for the bias in temperature has a comparable or perhaps even larger impact than the use of azimuthal median densities. 

Finally, we also show the joint effect of correcting densities (through the use of azimuthal medians) and temperatures (imposing $S/B > 0.2$) on the entropy profiles (red in Fig. \ref{ratio_flat}). To first order, this is the net sum of the previous two measurements. We notice a radial increase of the median ratio (reaching $\sim 1.13$ at $R_{500}$), meaning that the contribution from the systematics is larger in the outskirts than in the central regions. If these effects were not taken into account, they would have significant impact on the reconstruction of the entropy profiles, thus leading to incorrect conclusions on their shape in the cluster outskirts.

\subsection{Shape of the profiles}
\label{shape_profiles}

In this section, we present the results of the analytical fit to the individual HIGHMz entropy profiles using a constant plus power-law model of the form $K = K_0 + K_1(r/100~\mathrm{kpc})^\alpha$ \citep{donahue05}. This exercise allowed us to discuss both the central entropy ($K_0$) distribution, a topic that has been debated in the past decades, the distribution of the outer slopes $\alpha$, and to compare with results from previous works (\citeauthor{cavagnolo09} \citeyear{cavagnolo09}; \citeauthor{pratt10} \citeyear{pratt10}). 

\begin{figure*}
\centering
\sidecaption
            \includegraphics[width=0.4\textwidth]{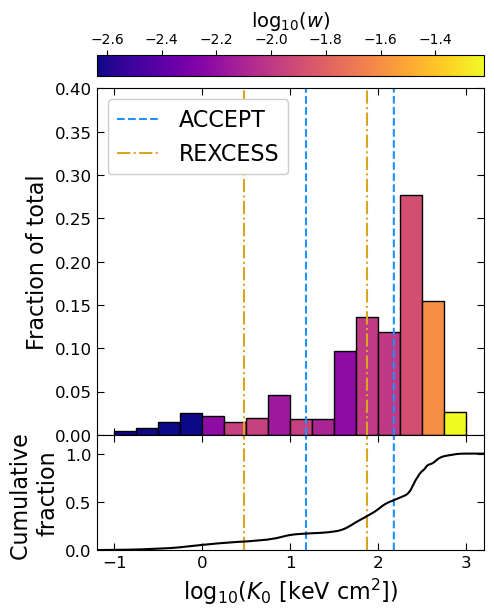}\hspace{0.7cm}
            \includegraphics[width=0.4\textwidth]{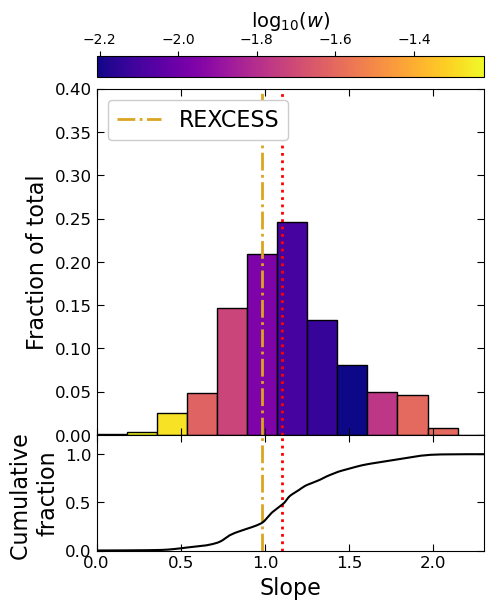}
            \caption{\footnotesize Distribution of central entropies $K_0$ (left) and outer slopes $\alpha$ (right) of the HIGHMz clusters. Plot bars are colour coded according to the median centroid shift $w$ in each bin. Bottom panels are the cumulative distributions of the two histrograms. Blue and golden vertical lines mark the positions of the peaks identified with ACCEPT \citep{cavagnolo09} and \rexcess\ \citep{pratt10}, respectively. The red dotted line on the right is the canonical slope value of $1.1$.}
            \label{K0_hist}
\end{figure*}

We performed the fits in a Bayesian framework using PyMC, assuming flat priors in a wide range for the three parameters to be determined. The best-fitting parameters for each cluster are given in Table \ref{table:singlean_fit}. To compare fairly with \citet{cavagnolo09} and \citet{pratt10}, we used the HIGHMz entropy profiles derived from mean densities (see Appendix \ref{sect_mean_entropy}). The histograms of the resulting central entropies $K_0$ and outer slopes $\alpha$ are shown in Fig. \ref{K0_hist} (top left and top right), together with the cumulative distributions (bottom panels), which are independent of the adopted binning. In constructing the histograms, we took into account the statistical uncertainties on the best-fitting parameters, in order to correctly weight each measurement. More specifically, for each measurement of $K_0$ and $\alpha$, we generated 1000 random realisations, normally distributed around the best-fitting value and using the statistical error as sigma, and then computed the histograms from these distributions, normalised to the total number of realisations. The histograms in Fig. \ref{K0_hist} are colour coded according to the median centroid shift of the clusters in each bin, in order to link the shape of the profiles to their morphological state.

\begin{table}
\caption{\footnotesize Best-fitting parameters of the constant plus power-law functional form describing the shape of HIGHMz entropy profiles.}
\centering    
\tiny
\begin{tabular}{cccc}    
\toprule
\toprule
Cluster & $K_0$ (keV cm$^2$)& $K_1$ (keV cm$^2$)& $\alpha$\\
& $U(0,1000)$ & $U(0,500)$ & $U(0,2)$ \\
\midrule         
PSZ2G004.45$-$19.55 & $41.78 \pm 17.78$ & $140.49 \pm 17.90$ & $1.10 \pm 0.07$ \\
PSZ2G008.94$-$81.22 & $48.42 \pm 38.20$ & $318.76 \pm 33.64$ & $0.47 \pm 0.03$ \\
PSZ2G044.77$-$51.30 & $99.44 \pm 30.77$ & $143.63 \pm 26.61$ & $0.79 \pm 0.08$ \\
PSZ2G046.10$+$27.18 & $449.65 \pm 27.28$ & $12.25 \pm 8.54$ & $1.56 \pm 0.28$ \\
PSZ2G056.93$-$55.08 & $345.95 \pm 10.90$ & $12.10 \pm 2.29$ & $1.88 \pm 0.08$ \\
PSZ2G057.25$-$45.34 & $54.06 \pm 11.94$ & $125.02 \pm 11.83$ & $1.11 \pm 0.05$ \\
PSZ2G072.62$+$41.46 & $245.74 \pm 13.36$ & $42.91 \pm 6.22$ & $1.53 \pm 0.07$ \\
PSZ2G073.97$-$27.82 & $7.04 \pm 4.81$ & $162.43 \pm 8.10$ & $1.06 \pm 0.03$ \\
PSZ2G092.71$+$73.46 & $74.28 \pm 31.59$ & $172.29 \pm 26.17$ & $0.80 \pm 0.06$ \\
PSZ2G111.61$-$45.71 & $260.46 \pm 23.64$ & $40.73 \pm 10.93$ & $1.44 \pm 0.12$\\ 
PSZ2G155.27$-$68.42 & $166.41 \pm 31.87$ & $87.51 \pm 22.49$ & $1.17 \pm 0.12$ \\
PSZ2G159.91$-$73.50 & $75.69 \pm 15.20$ & $127.11 \pm 13.64$ & $0.98 \pm 0.05$ \\
PSZ2G186.37$+$37.26 & $235.88 \pm 16.50$ & $64.01 \pm 9.27$ & $1.36 \pm 0.07$ \\
PSZ2G195.75$-$24.32 & $257.79 \pm 47.35$ & $70.61 \pm 27.08$ & $1.15 \pm 0.15$ \\
PSZ2G201.50$-$27.31 & $101.20 \pm 33.21$ & $123.27 \pm 26.56$ & $0.95 \pm 0.10$ \\
PSZ2G205.93$-$39.46 & $0.80 \pm 0.94$ & $132.59 \pm 3.64$ & $1.00 \pm 0.02$ \\
PSZ2G210.64$+$17.09 & $99.88 \pm 55.06$ & $135.03 \pm 45.82$ & $0.89 \pm 0.15$\\ 
PSZ2G216.62$+$47.00 & $38.14 \pm 26.70$ & $219.41 \pm 24.99$ & $0.72 \pm 0.05$ \\
PSZ2G228.16$+$75.20 & $225.16 \pm 33.93$ & $74.10 \pm 20.28$ & $1.13 \pm 0.11$ \\
PSZ2G239.27$-$26.01 & $304.54 \pm 24.95$ & $29.77 \pm 10.69$ & $1.64 \pm 0.18$ \\
PSZ2G243.15$-$73.84 & $249.59 \pm 30.27$ & $56.90 \pm 17.57$ & $1.25 \pm 0.14$ \\
PSZ2G262.27$-$35.38 & $223.47 \pm 98.85$ & $172.42 \pm 67.52$ & $0.79 \pm 0.13$ \\
PSZ2G266.04$-$21.25 & $283.26 \pm 16.02$ & $24.46 \pm 5.70$ & $1.74 \pm 0.11$ \\
PSZ2G277.76$-$51.74 & $480.74 \pm 60.76$ & $31.79 \pm 28.74$ & $1.26 \pm 0.32$ \\
PSZ2G278.58$+$39.16 & $3.10 \pm 3.59$ & $208.92 \pm 7.68$ & $0.76 \pm 0.02$ \\
PSZ2G284.41$+$52.45 & $44.98 \pm 6.57$ & $139.72 \pm 6.99$ & $1.13 \pm 0.03$ \\
PSZ2G286.98$+$32.90 & $372.00 \pm 56.08$ & $87.27 \pm 28.27$ & $1.27 \pm 0.14$ \\
PSZ2G324.04$+$48.79 & $0.62 \pm 0.72$ & $168.86 \pm 4.34$ & $1.03 \pm 0.02$ \\
PSZ2G340.36$+$60.58 & $0.37 \pm 0.43$ & $129.62 \pm 1.75$ & $1.13 \pm 0.01$ \\
PSZ2G340.94$+$35.07 & $7.79 \pm 1.97$ & $116.67 \pm 3.97$ & $1.21 \pm 0.02$ \\
PSZ2G346.61$+$35.06 & $638.96 \pm 148.32$ & $70.26 \pm 116.83$ & $0.79 \pm 0.47$\\ 
PSZ2G349.46$-$59.95 & $104.37 \pm 10.60$ & $79.00 \pm 8.17$ & $1.36 \pm 0.05$ \\
\bottomrule
\end{tabular}
\label{table:singlean_fit}
\end{table}

\subsubsection{Central entropy distribution}
\label{central_distr}

Based on an archival collection of 239 clusters observed with $Chandra$, known as ACCEPT, \citet{cavagnolo09} found a bimodal distribution of the central entropies $K_0$, with two peaks of similar amplitude located at $K_0 \sim 15$ keV cm$^2$ and $K_0 \sim 150$ keV cm$^2$. Conversely, by studying \rexcess\ clusters, \citet{pratt10} identified two tentative peaks at lower entropy ($K_0\sim 3$ and $\sim 75$ keV cm$^2$), with an amplitude ratio of $1:3$. However, they could not statistically distinguish between a bimodal and a left-skewed distribution of $K_0$. \citet{pratt10} proposed various reasons for the differences observed, including a possible overestimation of the central temperature distribution by \citet{cavagnolo09} due to the lack of deprojection of their measured 2D temperatures. Moreover, since ACCEPT is an archive limited sample, the clear gap identified by \citet{cavagnolo09} between $K_0 \sim 30-50$ keV cm$^2$ may be the simple result of scientists' prevailing interest in either strong cool-cores clusters or in energetic mergers, leaving out clusters with `average' properties, which will be observed in a representative sample like \rexcess. In summary, what the real distribution of central entropies in the cluster population is is still under debate. Clearly, given its close connection to the underlying physical mechanisms operating in cluster cores, the investigation of the central entropy distribution is a key issue for the cluster community.

We further investigated the topic of cluster central entropy distribution with the HIGHMz clusters. The results are shown in Fig. \ref{K0_hist} (left). Unlike \citet{cavagnolo09}, and more similarly to \citet{pratt10}, we do not find evidence for a bimodal distribution of central entropies, but rather for a left-skewed distribution, which peaks at $K_0 \sim 200$ keV cm$^2$, with an extended tail towards low $K_0$ values. This is not an artefact of the adopted binning, as supported by the cumulative distribution in the bottom panel of Fig. \ref{K0_hist} (left), which shows a gradual increase with $K_0$. This finding indicates that we observe more clusters with high central entropy in the present sample than in \rexcess\ and ACCEPT. Similar behaviour has been noted in Sect. \ref{other_samples} in comparison to X-COP and ESZ, where the median entropy profile of HIGHMz has been found to be flatter in the central regions (Fig. \ref{med_samples}). The small peak observed at $\sim 7$ keV cm$^2$ is simply the result of two clusters with similar central entropy (PSZ2G073.97$-$27.82 and PSZ2G340.94$+$35.07). We note that we restricted the histogram to values of $K > 0.1$ keV cm$^2$; however, four clusters have a best-fitting central entropy that is consistent with zero within $1 \sigma$.

Finally, we observe a clear correlation between $K_0$ and the centroid shift $w$, as indicated by the colour coding adopted in Fig. \ref{K0_hist}. Indeed, morphologically disturbed clusters tend to have higher central entropy, while morphologically relaxed clusters populate the left tail of the distribution, as already observed in Fig. \ref{K_prof}. Typically, $K_0$ is used to classify cool-core and non-cool-core clusters from a thermodynamical point of view; although it indicates the morphological state of a cluster as measured at large scales, $w$ shows a good correlation with $K_0$. As the median mass of HIGHMz is larger than both X-COP and ESZ, it may be that this very high mass population is dominated by disturbed objects with a high central entropy. Alternatively, the pure SZ-selection of HIGHMz may contribute to the presence of a greater number of disturbed, high central entropy systems \citep[][]{planck11_IX, rossetti16, rossetti17, andrade-santos17, lovisari17}.

Although our results seem to be in contrast with previous findings, they cannot be considered as definitive. In fact, the study of central regions is not straightforward, and the presence of additional uncertainties may bias our conclusions. For example, the physical size of the central bins is different for morphologically relaxed and disturbed clusters, a consequence of the adaptive binning method we used for the spectral extraction (\citeauthor{chen23} \citeyear{chen23}; see Sect. 3.2 of \citeauthor{rossetti24} \citeyear{rossetti24}), thus possibly altering to some level the measured distribution of the central entropies. In addition, differences in sample selection methods (X-rays and SZ), the adopted satellite for the study ($Chandra$ and XMM-$Newton$) and redshift ranges may prevent a fair comparison between the samples. Future studies on the entire CHEX-MATE sample will provide increased statistical power to derive additional constraints on the central entropy distribution in clusters. However, while CHEX-MATE was designed to ensure homogeneous observations at $R_{500}$, the same cannot be guaranteed for the central regions. Therefore, a dedicated program on a representative cluster sample, with the aim of reaching homogeneous spatial resolution in the central regions, may be needed to definitively shed light on this issue.

\subsubsection{External slopes of the profiles}
The right-hand panel of Fig. \ref{K0_hist} shows the distribution of outer slopes, $\alpha$, of the HIGHMz clusters, as from the constant plus power-law model. A wide range of values for the parameter $\alpha$ is observed, extending from $\sim0.5$ to $\sim 2$. Similarly to the findings by \citet{pratt10} for \rexcess\ clusters, there is no indication of bimodality in the distribution of slopes. The median of the measured distribution is $\alpha \sim 1.12$, remarkably close to the self-similar nominal value of $1.1$, and slightly larger than $0.98$, as measured by \citeauthor{pratt10} (\citeyear{pratt10}, shown by the golden vertical line in Fig. \ref{K0_hist} (right) using \rexcess. While the median slope of the \rexcess\ sample still falls within the large $1\sigma$ dispersion of HIGHMz ($\sigma = 0.35$), the lower median value likely reflects the presence of more lower mass systems in the \rexcess\ sample. Finally, the peak value measured by individual fits of the entropy profiles ($\alpha \sim 1.12$) is larger than the best fitting value presented in Fig. \ref{analytical_fit}, when all the HIGHMz entropy profiles are jointly fitted ($\alpha \sim 0.87$). Although the latter is still included within the $1\sigma$ distribution presented in this section, the observed differences may simply reflect the rigidity of the assumed model, which allows for larger variability when fitting individual profiles.

Similarly to the central entropies, the distribution of the outer slopes presented in Fig. \ref{K0_hist} (right) is also colour coded according to the median centroid shift parameter ($w$) of each bin. This allows us to link the measured outer slopes to the morphological state of the clusters. From the figure, it is evident that more relaxed clusters, with lower values of $w$, exhibit a slope $\alpha$ much closer to the canonical 1.1 value, with a relatively small scatter. Conversely, more disturbed clusters, with larger values of $w$, are more prevalent in the tails of the distribution, displaying both lower and higher values than $1.1$. This is also in line with the findings presented in Fig.~5 of \citet{pratt10}, showing that the best fitting slopes of the sub-sample of relaxed \rexcess\ clusters is closer to $\sim 1$ with a relatively small scatter, while dynamical disturbed objects are characterised by larger scatter in the $\alpha$ distribution. 


\subsection{Beyond the self similar scenario}
\label{Am_section}
In Sect. \ref{other_samples}, we compared entropy profiles of ESZ, X-COP, and \rexcess\ clusters to those from HIGHMz, noting similar shapes but significant differences for their normalisations. These discrepancies were observed to correlate with the median masses of each sample, with low mass clusters exhibiting more pronounced deviations. These findings support the idea that non-gravitational processes leave a signature in the physical properties of the ICM, particularly so for less massive systems, and so a self similar scaling (Eq. \ref{eq_k500}) may not be fully representative of all the physical processes affecting the gas entropy in a cluster.

Some works already focused on alternative rescalings of the thermodynamic profiles, that go beyond the self similar scenario and account for the effects of non-gravitational processes. For example, \citet{pratt10} showed a mass-dependent excess of the gas entropy in \rexcess\ clusters, while \citet{pratt22} highlighted a stronger than self similar evolution with redshift, together with a significant residual dependence on mass, for the density profiles of $\sim 120$ galaxy clusters. In the following, we investigate further the impact of non-gravitational processes through the combined sample of \rexcess, X-COP, ESZ, and HIGHMz clusters, to explore the possibility of mass and redshift dependencies of the gas entropy which go beyond the self similar scenario. 

We initially studied the impact of both an additional mass dependence and a modified evolution with respect to self-similar expectations. This investigation is detailed in Appendix \ref{App_Az_dependence}, where we show that we did not find strong statistical evidence for a modified redshift dependence in the scaling, likely due to the limited redshift range of the sample under consideration. We therefore assumed self-similar evolution and studied the impact of a residual mass dependence only. This is detailed in Sects. \ref{sect_global_Am} and \ref{sect_radial_Am}, where we investigate both global and radial effects, respectively. Finally, although we used the same scaling and adopted the same code to compute the medians of the different samples, some additional systematic effects, for example due to different data analysis approaches, may affect the following results to some degree. Through the future study on the entire CHEX-MATE sample, we will be able to derive definitive constraints on the departure from the self-similar scenario, minimising potential systematics through a homogeneous analysis of all the 118 clusters in the sample.
\begin{figure*}
\centering
            \includegraphics[width=0.42\textwidth]{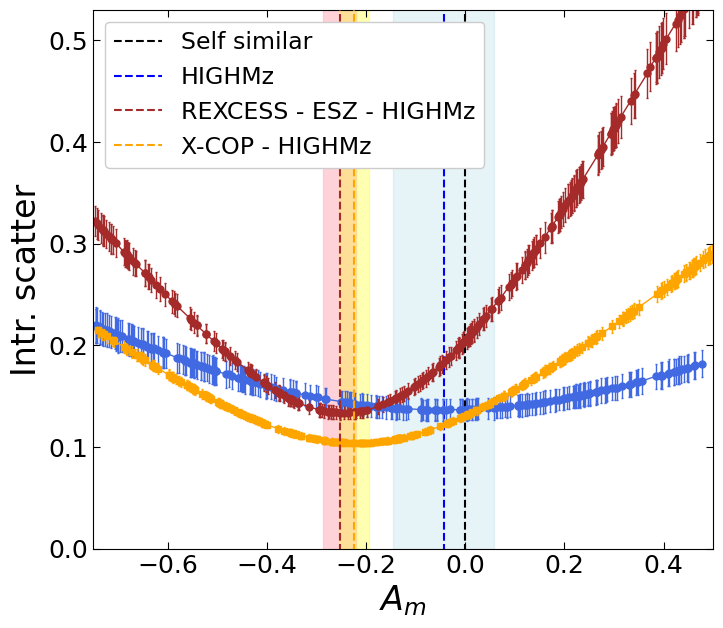}\hspace{0.5cm}
            \includegraphics[width=0.42\textwidth]{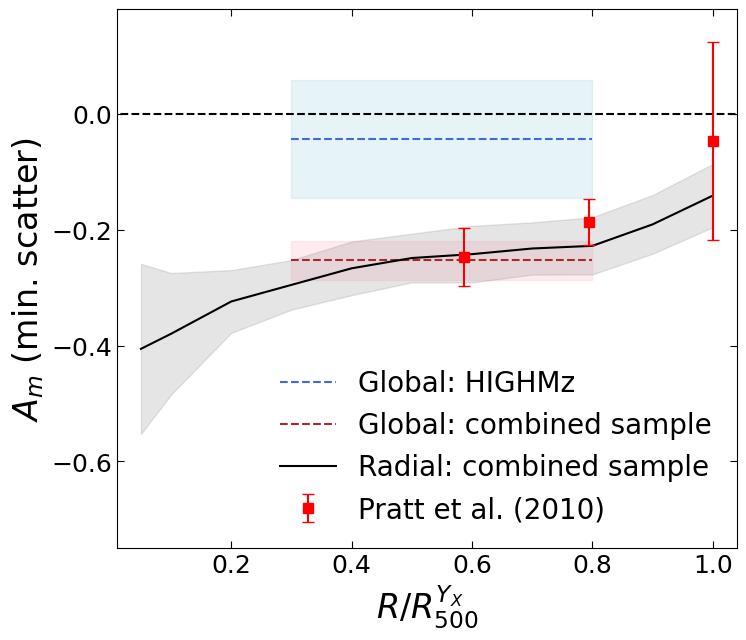}
            \caption{\footnotesize Constraints to the residual mass dependence $A_m$ in the entropy rescaling. Left panel: distribution of the intrinsic scatter of different samples as a function of $A_m$, to identify the global (i.e. within [$0.3-0.8$]~$R_{500}$) $A_m$ correction (Sect. \ref{sect_global_Am}). Yellow is used for the combined X-COP -- HIGHMz sample, brown for \rexcess\ -- ESZ -- HIGHMz, while blue for HIGHMz only. Vertical dashed lines and shaded areas mark the positions of the minima, together with their statistical errors. Right panel: radial dependence of the $A_m$ values which minimise the scatter of the combined \rexcess~-- ESZ -- HIGHMz sample (Sect. \ref{sect_radial_Am}), together with their statistical errors (black solid line and grey shaded area). Shown in brown (blue) is the global measurements using \rexcess\ -- ESZ -- HIGHMz (HIGHMz only) sample. In red are results from \rexcess\ only (\citeauthor{pratt10} \citeyear{pratt10}). Both in the left and right panels, black dashed lines are the prediction from self-similar scenario, i.e. $A_m = 0$.}
            \label{Am}
\end{figure*}

\subsubsection{Global dependence on mass}
\label{sect_global_Am}
We assumed that the differences between the median entropy profiles of the four samples can be entirely explained by a modified dependence on cluster mass, rather than a self similar one. We thus introduced the parameter $A_m$, that quantifies the departure from self similar predictions, and built a modified entropy rescaling:
\begin{equation}
    \widetilde{K_{500}} = M_{500}^{2/3 + A_m}~E(z)^{-2/3}.
    \label{eq_am}
\end{equation}
To identify the value of $A_m$ that is needed to explain the differences we have observed in Sect. \ref{other_samples}, we studied how the scatter of the measured profiles varies as a function of $A_m$. Our best-fitting residual mass dependence is then the value of $A_m$ that minimises the dispersion of the profiles in a given radial range.

To do so, we followed a procedure similar to the one described in Sect. \ref{codes_median}. Specifically, we fitted together the entropy profiles of the considered samples in the radial range $[0.3-0.8]~R_{500}$, thus excluding the flattening that is observed in the core region and potentially biased measurements in the outskirts, using a power law and an intrinsic scatter:
\begin{equation}
    K(x)/\widetilde{K_{500}}=A\cdot x^\alpha \cdot \text{exp}(\pm \sigma_{\mathrm{int}}),
    \label{pow_eq}
\end{equation}
where again $x = R/R_{500}$. To investigate the dependence on the residual mass dependence, we generated $250$ random values of $A_m$, uniformly distributed in the range $[-0.75,+0.5]$. At each iteration, the entropy profiles were rescaled using a different realisation of $\widetilde{K_{500}}$ and the best-fitting intrinsic scatter was then plotted as a function of $A_m$. Similarly to what was discussed in Sect. \ref{other_samples}, we specify that we considered X-COP and HIGHMz and \rexcess, ESZ, and HIGHMz separately in the analysis, to reduce the impact of systematic errors coming from a heterogeneous cluster sample. We also noticed that some clusters are in common between HIGHMz, ESZ, and \rexcess; we thus excised duplicates that may bias our results, keeping our measurements first and then, if a cluster was only in common between \rexcess\ and ESZ, those reported in \citet{pratt10}. The final combined samples count 44 clusters for the combined X-COP -- HIGHMz sample, and 108 for \rexcess\ -- ESZ -- HIGHMz. 

The results are presented in Fig. \ref{Am} (left), where we plot the distribution of the measured dispersions as a function of the residual mass dependence $A_m$. We immediately notice that these describe well defined curves in the plane $A_m-\sigma_{\mathrm{int}}$ and that the minima of these curves are not achieved at $A_m=0$, as would be the case if cluster entropies scaled self similarly. The exact positions of the minima and the associated errors were identified through a Monte Carlo simulation; specifically, i) we generated 1000 random realisations of each data point around its statistical error; ii) each time we linearly interpolated the randomly generated curve and found its minimum; and iii) we computed the median and $1\sigma$ dispersion from this distribution. The measured errors give an idea of the flattening of each curve around the position of the minimum. We measure $A_m$ (min. scatter) $= -0.24 \pm 0.03$ and $-0.21 \pm 0.04$ for the combined samples of \rexcess\ -- ESZ -- HIGHMz (brown in Fig. \ref{Am}, left) and X-COP -- HIGHMz (yellow), respectively. We thus find evidence that a milder mass dependence than self similar is needed to reduce the scatter of the measured entropy profiles to its minimum in the $[0.3-0.8]~R_{500}$ radial range, supporting the idea that non-gravitational processes break the self similarity of the observed clusters and cause an enhancement of their gas entropy that depends on cluster mass. We note that the measured values of $A_m$ (min. scatter) for the two combined samples are in good agreement within their error bars. This is indeed expected, as \rexcess\ and ESZ together cover the mass range of the X-COP sample. 

Finally, we also repeated the same exercise using the HIGHMz clusters only (blue in Fig. \ref{Am}, left), to test whether the departure from the self similar scaling is observed also at the massive cluster scale. We find $A_m$ (min. scatter) $=-0.04 \pm 0.10$, consistent with zero within the large error bar. This supports the idea that, beyond the core, the properties of massive clusters are mainly driven by the gravitational assembly, which results in a good agreement with the self-similar predictions for the rescaling. 

\subsubsection{Radial dependence on mass correction}
\label{sect_radial_Am}
In the previous section, we identified a value $A_m$ that would explain the global differences observed between samples at different masses. However, \citet{pratt10} also identified a radial trend of the mass-dependent entropy excess, that is stronger at $\sim R_{2500}$ and becomes negligible at $\sim R_{500}$ within the large observational uncertainties. The increased sample size of the present study allows to conduct a similar investigation with improved precision, and so to map a radial dependence of the measured $A_m$ values that minimise the dispersion of the profiles. To do so, we repeated the procedure presented in Sect. \ref{sect_global_Am} in a slightly different fashion. Specifically, we first interpolated the profiles on a common grid of 10 equally-spaced radial bins within the range $[0.1-1]~R_{500}$ (plus an additional measurement at $0.05~R_{500}$), similarly to Sect. \ref{other_samples}. We then fitted the data points belonging to each radial bin with the model: 
\begin{equation}
    K/\widetilde{K_{500}}=A\cdot \text{exp}(\pm \sigma_{\mathrm{int}}),
\end{equation}
and studied the distribution of the intrinsic scatter as a function of $A_m$ at each radius. Similarly to what was done in Sect. \ref{sect_global_Am}, we generated $100$ random values of $A_m$ to use in the rescaling, uniformly distributed in the range $[-0.75,0.25]$. However, for this test we used only the combined sample of \rexcess\ -- ESZ -- HIGHMz, since the larger number of objects affords better precision on the radial dependence of the mass correction. The position of the minimum value of each distribution and the associated errors were computed as in Sect. \ref{sect_global_Am}. These are shown in Fig. \ref{Am} (right), as a function of radius and compared to previous measurements by \citet{pratt10}.

We find a radial trend of the $A_m$ values minimising the ten distributions, meaning that the minimum scatter is achieved at lower $A_m$ values at small radii, while a weaker correction is needed moving towards the cluster outskirts. More specifically, we measure $A_m$ (min. scatter) $= -0.41 \pm 0.15$ at $0.05 ~R_{500}^{Y_{\rm X}}$, while this value is closer to self similar predictions at $R_{500}^{Y_{\rm X}}$, where we find $A_m$ (min. scatter) $= -0.14 \pm 0.06$. Our measurements are in agreement with previous findings by \citet{pratt10} and support the idea that non-gravitational processes have larger effects on the ICM properties at small radii, while corrections to self similar predictions are smaller moving towards the outskirts. Interestingly, at $R_{500}^{Y_{\rm X}}$ their measurement is consistent with zero within the large error bars; with increased statistics, we are able to constrain a residual mass dependence other than zero also at this external radius. 

Our measured value of $A_m$ (min. scatter) at $R_{500}^{Y_{\rm X}}$ is in agreement with predictions based on the semi-analytic model proposed by \citet{ettori23},  i.e. $A_m$ (min. scatter) $ = -0.16 \pm 0.01$, although they also considered a modified dependence on redshift, rather than assuming a self similar evolution as we have done. At the same radius, our measurement is also consistent within $1\sigma$ with results from simulations implementing full baryonic physics, as presented in \citet{nagai07}, which predict $A_m = -0.062 \pm 0.036$.


\subsection{Rescaling using weak lensing masses}
\label{scal_WL}

In Sect. \ref{sec_ent_results}, we have adopted masses derived from $Y_X$ to build the rescaling of the measured entropy profiles, due to their lower sensitivity to cluster dynamical state compared to hydrostatic equilibrium masses. However, \citet{pratt22} have shown that the use of $M_{500}^{Y_{\rm X}}$ leads to a systematic suppression of the true dispersion of the density profiles, for a sample of $\sim$120 clusters. This happens as a cluster with density higher than the average (at a fixed mass) will possess greater gas mass, resulting in a higher measured $Y_X$ value. Consequently, this will lead to an overestimate of $M_{500}^{Y_{\rm X}}$ (and so $R_{500}^{Y_{\rm X}}$) that, when used in the rescaling, will move the measured density profile back to the mean one, thus reducing the apparent scatter. In other words, \citet{pratt22} showed that the covariance between $M_{500}^{Y_{\rm X}}$ and gas density implies a suppression of the intrinsic scatter in rescaled density profiles.

In this section, we investigate further the same issue using the entropy profiles of HIGHMz. In particular, to test the degree of covariance between our profiles and $M_{500}^{Y_{\rm X}}$, we performed an alternative rescaling, using masses derived from a ground-based weak lensing analysis (hereafter $M_{500}^{\rm WL}$). These masses are not estimated from any of the ICM observables and do not rely on any assumptions about the dynamical state of the clusters; therefore they are commonly considered a good estimate of the `true' mass of a cluster in terms of accuracy (\citeauthor{becker11} \citeyear{becker11}; \citeauthor{pratt19} \citeyear{pratt19}; \citeauthor{umetsu20_rev} \citeyear{umetsu20_rev}; \citeauthor{euclid_giocoli24} \citeyear{euclid_giocoli24}). However, an unbiased reconstruction of the weak-lensing signal (or reduced tangential shear profile) within the cluster extension is not straightforward and is particularly challenging for individual low-mass systems ($M_{500}\sim 10^{14}$ M$_{\odot}$; see Fig. 17 of \citeauthor{umetsu20} \citeyear{umetsu20}) due to their shallower potential well, and for distant ones, as it is more difficult to have information on a sufficient number of background galaxies. As a result, weak lensing masses are often accompanied with large statistical errors or, in other terms, they are nearly unbiased, but not precise.

\begin{figure}[h!]
\centering
\includegraphics[width=0.44\textwidth]{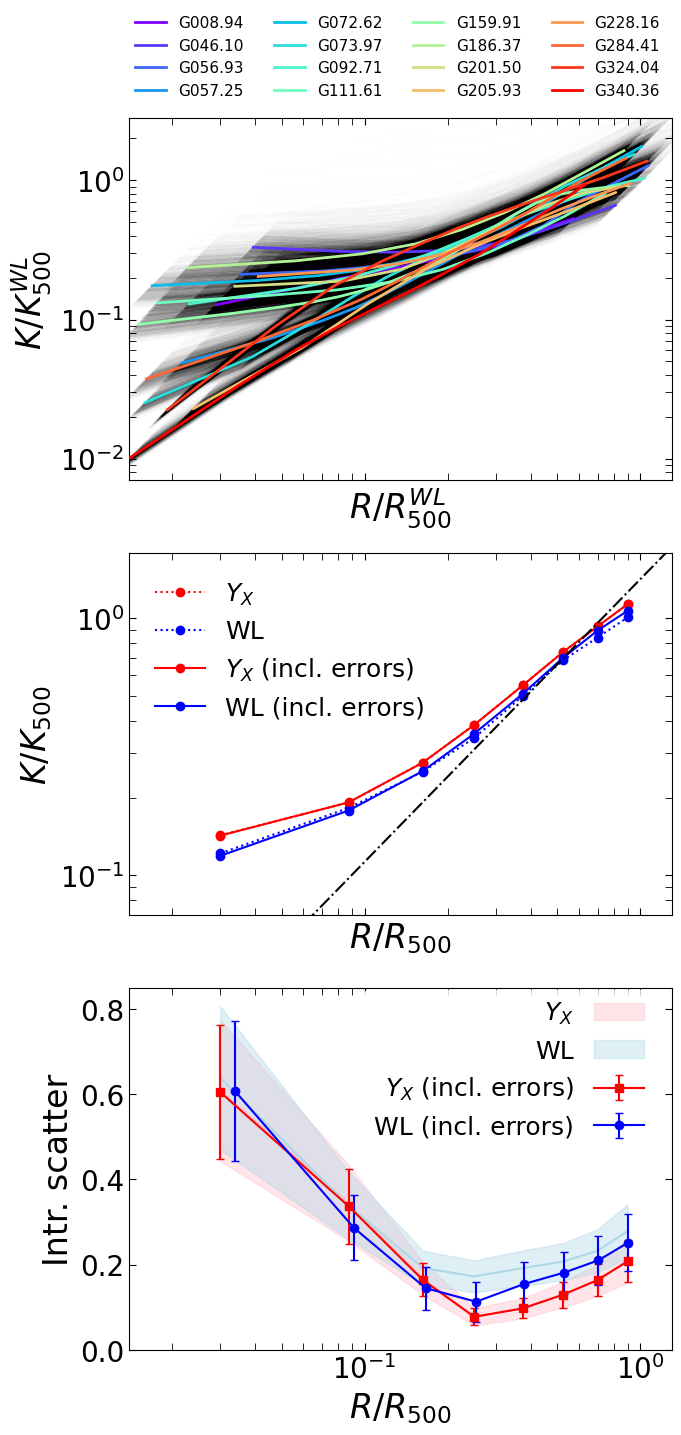}
            \caption{\footnotesize Rescaling using weak-lensing masses. From top to the bottom: i) entropy profiles of the sub-sample of 16 clusters with weak lensing masses available, scaled using $M_{500}^{\rm WL}$, where shaded grey lines show the impact of accounting for statistical errors on $M_{500}^{\rm WL}$ in the rescaling, as described in the text; ii) median entropy profiles, obtained using $M_{500}^{Y_{\rm X}}$ (red dotted line) and $M_{500}^{\rm WL}$ (blue dotted line) in the rescaling; continuous lines are the measurements after including statistical errors on mass measurements; iii) same as centre, but for the intrinsic dispersion; here, simple scalings without including errors on mass measurements are shown as shaded areas.}
            \label{WL_scal}
\end{figure}

For our purpose, we took advantage of weak-lensing $M_{500}$ mass estimates available in the Amalgam2 dataset based on archival Subaru Suprime-Cam and CFHT MegaPrime/MegaCam observations (Umetsu et al. and Gavazzi et al., in preparation), produced within the CHEX-MATE collaboration. These weak-lensing masses were inferred from the reduced tangential shear profiles of individual clusters assuming the spherical NFW profile \citep{NFW96} with uninformative priors on the halo mass and concentration ($M_{200}$, $c_{200}$), closely following the procedures outlined in \citet{umetsu20}. To date, these masses have been measured for 42 CHEX-MATE clusters including a subset of HIGHMz comprising 16 clusters, that we then considered for the present study; the entropy profiles of this sub-sample, scaled using $M_{500}^{\rm WL}$, are illustrated in Fig. \ref{WL_scal} (top). 

To test the impact of the adopted mass in the rescaling on the measured dispersion of the profiles, we employed the same fitting procedure outlined in Sect. \ref{sec_ent_results}. Specifically, we adopted the model described in Eq. \ref{eq1}, both considering $M_{500}^{Y_{\rm X}}$ and $M_{500}^{\rm WL}$ to build $K_{500}$ and $R_{500}$, to test the impact of the two mass rescalings on the measured intrinsic scatter. 

The measured medians with this first fitting procedure are shown in Fig. \ref{WL_scal} (centre), as dotted red ($M_{500}^{Y_{\rm X}}$) and blue ($M_{500}^{\rm WL}$) lines, while the intrinsic scatter profiles are plotted in the bottom panel, as shaded pink and light blue envelopes, respectively. The figure shows that using weak lensing masses leads to a systematic decrease (a flat $\sim 10\%$) of the median profile at all radii and a net increase ($\gtrsim 50\%$) of the measured dispersion beyond $0.3~R_{500}$. However, this might not be the result of the covariance between the entropy profiles and $M_{500}^{Y_{\rm X}}$, but it may simply reflect the impact of the larger statistical errors associated to the $M_{500}^{\rm WL}$ measurements. In fact, the median fractional statistical uncertainty on weak lensing masses is $\sim 28\%$\footnote{Here, the weak-lensing analysis accounts for the shape noise and the cosmic noise due to uncorrelated large scale structure projected along the line of sight (see \citeauthor{umetsu20} \citeyear{umetsu20}).}, while that of masses derived from $Y_X$ is  $\sim 3\%$.

To explore their impact on the final measurements, we propagated the statistical errors associated to mass measurements into the scaled entropy profiles. More specifically, we generated 1000 random realisations of $M_{500}^{Y_{\rm X}}$ and $M_{500}^{\rm WL}$ for each cluster, normally distributed around the measured values using the $68\%$ dispersion as sigma, that we used to scale the profiles. Figure \ref{WL_scal} (top) also illustrates the spread around the measured entropy profiles due to mass randomisation, when scaling using $M_{500}^{\rm WL}$. From the random distribution around each entropy bin, we computed (as $16^{\text{th}}$ and $84^{\text{th}}$ percentiles) a new statistical error associated to each measurement, that quantifies the impact of mass randomisation. This additional error was then used to build an updated version of the error measurements at each bin:
\begin{equation}
    \sigma_{\mathrm{incl.~errors}} = \sqrt{\sigma_{\mathrm{stat.}}^2 + \sigma_{\mathrm{ran.~masses}}^2},
\end{equation}
including the original measured statistical errors ($\sigma_{\mathrm{stat.}}$) and the spread due to mass randomisation ($\sigma_{\mathrm{ran.~masses}}$). Adopting Eq. \ref{eq1}, we then recalculated the median and the intrinsic scatter of the scaled profiles, this time including the information from mass randomisation. The results of these new fitting procedures are shown in Fig. \ref{WL_scal} (centre and bottom) as red ($M_{500}^{Y_{\rm X}}$) and blue ($M_{500}^{\rm WL}$) solid lines. 

Accounting for the statistical errors associated to weak lensing mass measurements has small impact on the measured median (lower than $5\%$ at all radii), while it systematically reduces the dispersion (average $\sim 15\%$ beyond the core), reflecting the impact of their large uncertainties. Conversely, the small errors on $M_{500}^{Y_{\rm X}}$ barely affect the median and intrinsic scatter when employing this mass rescaling. A comparison between the two rescalings suggests that employing $Y_X$ masses leads to a $\sim 30\%$ suppression of the intrinsic scatter of the profiles beyond $\sim 0.2 ~R_{500}$,  although this result is not statistically significant. In the core, no evident differences are found. This would point to the conclusion that a degree of covariance of entropies and $Y_X$ is indeed observed, and using weak lensing masses would allow us to recover the `true' entropy scaling. As already mentioned, a similar behaviour was also observed by \citet{pratt22}; assuming that the suppression profile of their Fig.~10 is translated into a similar one for entropies, we would find a correction of $\sim 35 \%$ beyond the core, similar to the one reported here. 

However, the limited size of the sample adopted for the present study precludes definitive conclusions on this issue. The future study on the entropy profiles for the full CHEX-MATE sample, together with a similar investigation performed on the other thermodynamic quantities, will allow more robust conclusions on the impact of different mass measurements on the reconstruction of the intrinsic scatter of the thermodynamic profiles in galaxy clusters.  


\section{Summary and conclusions}
\label{summary}
The present study focused on the entropy profiles of 32 galaxy clusters, with masses $M_{500}^{Y_{\rm SZ}} > 7.75 \times 10^{14}$ M$_{\odot}$ and located at redshifts $0.2 < z < 0.6$. This sample, named HIGHMz, was selected from the parent CHEX-MATE sample \citep{chexmate21} to be representative of the most massive cluster population. The high masses of HIGHMz clusters allowed us to investigate the cluster entropy in the gravity-dominated regime, where non-gravitational processes are expected to have only minor effects, thus setting a valuable baseline for future studies at lower mass scales and higher redshifts. The ultimate goal of this work is to provide a set of observational constraints on the entropy profiles of the most massive clusters of the Universe, which will serve as a future reference for cosmological simulations and theoretical models describing entropy production and modification mechanisms at the galaxy cluster scale.

In general, the entropy profiles of HIGHMz clusters share similar properties to those measured in previous studies (e.g. \citeauthor{cavagnolo09} \citeyear{cavagnolo09}; \citeauthor{pratt10} \citeyear{pratt10}; \citeauthor{ghirardini19} \citeyear{ghirardini19}). For example, we found that:
\begin{itemize}
    \item notable dispersion is present in the core regions, while the profiles scale more self similarly and approach predictions from non-radiative simulations \citep{voit+05} towards the cluster outskirts;
    \item the central dispersion reflects the morphological variety of the clusters in the sample, as revealed by the strong correlation with morphological indicators;
    \item compared with the predictions from non-radiative simulations, beyond the core the profiles are well fitted by a power law with gentler slope at all radii ($\alpha \sim 0.9$ at $R_{500}$; Fig. \ref{analytical_fit}), leading to a lower ($\sim 10\%$) measured entropy in the cluster outskirts. 
\end{itemize} 
In contrast with \citet{cavagnolo09} and more similarly to \citet{pratt10}, we found no evidence of bimodality in the central entropy distribution (Fig. \ref{K0_hist}).

We investigated the relative role of possible systematic errors in the entropy reconstruction. We showed that:
\begin{itemize}
    \item the exclusion of temperature measurements in radial bins with low source-to-background ($S/B < 0.2$, \citealt{rossetti24}) from the deprojection, which lie mostly beyond $R_{500}$, allows us to regularise the entropy profiles in the outskirts. The change in the external part of the profiles is larger than 10\% only for four clusters, which are those featuring a very low temperature in the last radial bin (Fig. \ref{2D_prof}).  After exclusion of these temperature measurements, these profiles more closely approach the predictions from non-radiative simulations;
    \item the use of azimuthal median densities, which allows a more efficient excision of dense clumps, impacts the final shape of the entropy profiles at all radii (Fig. \ref{ratio_flat}), but in the cluster outskirts the temperature measurements are the major contributor to the systematic uncertainties.
\end{itemize}

We compared our profiles with other observational samples, namely \rexcess\ \citep{pratt10}, ESZ \citep{planck11_ESZ62}, and X-COP \citep{eckert22}. Entropy profiles are all in good agreement in shape, but some differences in normalisation are present, with the low-mass clusters showing higher scaled entropy at all radii (Fig. \ref{med_samples}). These discrepancies correlate with the median masses of the four samples and are a clear indication of the impact that non-gravitational processes have on the properties of the ICM. Through the combined sample of HIGHMz, ESZ, and \rexcess, we investigated the possibility of an alternative scaling of entropy with mass (Eq. \ref{eq_am}), that can account for the break of the self similarity due to non-gravitational processes. In particular, we found that:
\begin{itemize}
    \item a weaker mass dependence than self similar ($A_m \sim -0.25$) is able to minimise the dispersion of the entropy profiles in the radial range $[0.3-0.8]~R_{500}$ (Fig. \ref{Am}, left);
    \item the effect is radial (Fig. \ref{Am}, right), with stronger deviations from self similarity at small radii, although a non-zero mass dependence was measured also at $R_{500}$, where we found $A_m = -0.14 \pm 0.06$;
    \item the same study, performed on HIGHMz clusters only, showed that the $A_m$ value minimising the scatter is consistent with being zero, i.e. self similar;
    \item finally, we also investigated in Appendix \ref{App_Az_dependence} the possibility of a modified scaling also for the evolution with redshift (Eq. \ref{eq_az}), although the small considered redshift range did not allow definitive conclusions.
\end{itemize}
  
We also compared our results to recent simulations (The300, \citeauthor{cui18} \citeyear{cui18}; MACSIS, \citeauthor{barnes17} \citeyear{barnes17}) that are able to produce a sufficient number of massive clusters at redshifts $0.2 < z < 0.6$. We found that:
\begin{itemize}
    \item MACSIS clusters are in good agreement with observations, both in shape and in normalisation (Fig. \ref{K_sim}), where differences lower than $\sim 6\%$ are measured. Slightly larger differences (offset $\sim -9\%$) are observed in comparison with The300;
    \item regarding the correlation with morphological indicators, simulations are able to reproduce the measured behaviour for observations. We verified this both through sub-samples of defined relaxed and disturbed clusters (Fig. \ref{subsample_sim}) and, more quantitatively, through a correlation of the gas entropy with $c$ and $w$ as a function of radius (Fig. \ref{perason_corr}).
\end{itemize}

Finally, we tested the rescaling of the entropy profiles adopting masses derived from a weak lensing analysis and produced within the CHEX-MATE collaboration, in order to reduce covariance (as highlighted by \citeauthor{pratt22} \citeyear{pratt22} for densities) and to recover the `true' dispersion. We considered for this study a subset of 16 HIGHMz clusters, for which weak lensing masses were already available. We found that:
\begin{itemize}
    \item the different statistical errors on weak lensing and $Y_X$ masses have an impact on the results, if they are not properly taken into account in the analysis;
    \item after propagating the statistical errors on mass measurements, there is an indication that the rescaling through $Y_X$ masses leads to a systematic suppression ($\sim 30\%$, although not statistically significant) of the intrinsic dispersion of the profiles beyond the core (Fig. \ref{WL_scal}).
\end{itemize}

Although we managed to derive several observational constraints on the properties of the gas entropy at the massive cluster scale, we could not provide definitive conclusions on a few other aspects. For example, both the study of the central entropy distribution and the rescaling using weak lensing masses would require a larger cluster sample size. Furthermore, as discussed in Sect. \ref{other_samples}, we emphasise that in comparing with other observational samples we tried to minimise systematic errors that may result from a heterogeneous sample, for example by using the same code to compute median profiles and adopting the same mass rescalings. However, some residual hidden systematics may still be present, affecting our results to some degree. Further constraints on these and more aspects will be possible through the future study of the entropy profiles of the entire CHEX-MATE sample. 

In conclusion, our results have set an observational baseline for the radial entropy distribution in the highest mass systems, where gravitational infall is expected to be the dominant entropy generation mechanism. Outside the core regions, the profiles converge rapidly, both in slope and normalisation, to the theoretical expectation from gravity-only numerical simulations. The results further confirm the findings of \citet{pratt10}, in the sense that high central entropy is correlated with the morphological appearance of a cluster, and that low central entropy is correlated with the presence of a cool core. In this paper, for the first time, we have confirmed these same trends in recent cosmological numerical simulations of objects of similar mass. If ICM morphology is correlated to dynamical state \citep[e.g.][]{bauer05, poole06, jeltema08, chon12, rasia13, lovisari17, campitiello22}, these results therefore indicate that a high central entropy at these high masses is caused by gas mixing due to merging processes.

Comparison of the HIGHMz sample with other samples from the literature at lower average mass reveals the presence of additional entropy, above that expected from gravitational generation alone. A modification of the expected scaling to take this into account reveals a mass-dependence that is significantly less than self-similar. In other words, the ICM in lower mass systems has a higher entropy than that expected from gravitational collapse alone. Outside the core regions, this entropy increase is radially dependent, reaching larger radius in lower mass systems. This is consistent with a scenario where radiative cooling of the gas, or additional energy injection from  AGN activity or supernovae, raises the ICM in lower mass systems to a higher adiabat.

An important avenue for further work will therefore be to disentangle the non-gravitational energy injection from the effect of gas mixing due to merging activity. We intend to pursue this with the full CHEX-MATE sample.


\begin{acknowledgement}
    We thank the anonymous referee for useful comments, which have improved our paper. This work is based on observations obtained with XMM-\textit{Newton}, an ESA science mission with instruments and contributions directly funded by ESA Member States and NASA. The research has received funding from the European Union’s Horizon 2020 Programme under the AHEAD2020 project (grant agreement n. 871158). This research was supported by the International Space Science Institute (ISSI) in Bern, through ISSI International Team project \#565 “Multi-Wavelength Studies of the Culmination of Structure Formation in the Universe”. G.R. acknowledges C. Grillo for useful discussions. G.R., M.R., I.B., H.B., S.D.G., F.D.L., S.E., F.G., S.G., L.L., P.M., and S.M. acknowledge the financial contribution from the contracts Prin-MUR 2022, supported by Next Generation EU (n.20227RNLY3 {\it The concordance cosmological model: stress-tests with galaxy clusters}), ASI-INAF Athena 2019-27-HH.0, ``Attivit\`a di Studio per la comunit\`a scientifica di Astrofisica delle Alte Energie e Fisica Astroparticellare'' (Accordo Attuativo ASI-INAF n. 2017-14-H.0). G.W.P acknowledges long-term support from CNES, the French space agency. K.U. acknowledges support from the National Science and Technology Council of Taiwan (grant NSTC 112-2112-M-001-027-MY3) and the Academia Sinica Investigator Award (grant AS-IA-112-M04). L.L. also acknowledges support from INAF mini grant 1.05.12.04.01. M.S. acknowledges financial contributions from the contracts Prin-MUR 2022 supported by Next Generation EU (M4.C2.1.1, n.20227RNLY3 {\it The concordance cosmological model: stress-tests with galaxy clusters}) and INAF Theory Grant 2023: Gravitational lensing detection of matter distribution at galaxy cluster boundaries and beyond (1.05.23.06.17). H.B., P.M., and F.D.L. acknowledge financial contribution from the contracts ASI-INAF Athena 2019-27-HH.0, "Attività di Studio per la comunità scientifica di Astrofisica delle Alte Energie e Fisica Astroparticellare" (Accordo Attuativo ASI-INAF n. 2017-14- H.0), support from INFN through the InDark initiative, from “Tor Vergata” Grant “SUPERMASSIVE-Progetti Ricerca Scientifica di Ateneo 2021”, and from Fondazione ICSC, Spoke 3 Astrophysics and Cosmos Observations. National Recovery and Resilience Plan (Piano Nazionale di Ripresa e Resilienza, PNRR) Project ID CN00000013 ‘Italian Research Center on High Performance Computing, Big Data and Quantum Computing’ funded by MUR Missione 4 Componente 2 Investimento 1.4: Potenziamento strutture di ricerca e creazione di "campioni nazionali di R\&S (M4C2-19 )" - Next Generation EU (NGEU). H.B., P.M. and F.D.L. also acknowledge the financial contribution from the contract PrinMUR 2022 supported by Next Generation EU (n.20227RNLY3 The concordance cosmological model: stress-tests with galaxy clusters). Ma.G. acknowledges support from the ERC Consolidator Grant \textit{BlackHoleWeather} (101086804). B.J.M acknowledges support from Science and Technology Facilities Council grant ST/Y002008/1. E.P. acknowledges the support of the French Agence Nationale de la Recherche (ANR), under grant ANR-22-CE31-0010 (project BATMAN). J.S. was supported by NASA Astrophysics Data Analysis Program (ADAP) Grant 80NSSC21K1571. We made use of the following software packages: FTOOLS \citep{blackburn95}, DS9 \citep{joye03}, and several python libraries, such as numpy \citep{harris20}, matplotlib \citep{hunter07}, Astropy (\citeauthor{astropy13} \citeyear{astropy13}, \citeyear{astropy18}), PyMC \citep{abril-pla23}. 
\end{acknowledgement}


\bibliographystyle{aa} 
\bibliography{main} 

\begin{appendix}

\section{The300 and MACSIS datasets}
\label{App_sim}

We report in Tables \ref{table:the300} and \ref{table:MACSIS} additional information on the selected The300 and MACSIS clusters, respectively, which were used for comparison in Sect. \ref{simulations}. In particular, we specify their masses ($M_{500}$, and so the derived $R_{500}$), used in the rescaling of the profiles, together with morphological indicators such as $c$ and $w$, computed as described in \citet{campitiello22} and \citet{towler23}, respectively.

\begin{table}[h!]
\caption{\footnotesize The300 clusters considered for this study.}
\centering                         
\begin{tabular}{ccccc}        
\toprule
\toprule
Cluster & $M_{500}$  & $R_{500}$ & $c$ &$w$ \\
& $10^{14}$ M$_{\odot}$ & Mpc  & & \\
\midrule 
CL0002 &  19.965 & 1.740 & 0.34 & 0.014 \\
CL0006 &  11.303 & 1.439 & 0.50 & 0.004 \\
CL0009 &  14.291 & 1.556 & 0.43 & 0.016 \\
CL0011 &  12.193 & 1.476 & 0.18 & 0.040 \\
CL0017 &  10.215 & 1.391 & 0.42 & 0.003 \\ 
CL0018 &  11.297 & 1.439 & 0.39 & 0.007\\ 
CL0019 &  9.479 & 1.357 & 0.28 & 0.020 \\
CL0021 &  10.124 & 1.387 & 0.24 & 0.042 \\
CL0024 &  17.028 & 1.650 & 0.14 & 0.083 \\
CL0027 &  10.716 & 1.414 & 0.49 & 0.011 \\
CL0029 &  12.293 & 1.480 & 0.37 & 0.012 \\
CL0030 &  12.584 & 1.492 & 0.47 & 0.005 \\
CL0031 &  10.034 & 1.383 & 0.23 & 0.041 \\
CL0035 &  9.792 & 1.372 & 0.52 & 0.008 \\
CL0036 &  11.514 & 1.448 & 0.11 & 0.083\\
CL0039 &  11.031 & 1.428 & 0.43 & 0.012\\
CL0044 &  9.488 & 1.358 & 0.23 & 0.019 \\
CL0052 &  9.512 & 1.359 & 0.57 & 0.007 \\
CL0067 &  10.928 & 1.423 & 0.50 & 0.014\\
CL0082 &  11.617 & 1.452 & 0.38 & 0.061\\
CL0095 &  9.804 & 1.373 & 0.72 & 0.003\\
CL0097 &  9.339 & 1.350 & 0.23 & 0.063 \\
CL0130 &  10.654 & 1.411 & 0.20 & 0.067 \\
CL0141 &  10.052 & 1.384 & 0.38 & 0.006\\
CL0142 &  12.275 & 1.479 & 0.13 & 0.054\\
\bottomrule                             
\end{tabular}
\tablefoot{In the Table, we list $M_{500}$, $R_{500}$, and the morphological parameters $c$ and $w$, computed as described in \citet{campitiello22}.}             
\label{table:the300}
\end{table}

\begin{table*}
\caption{\footnotesize MACSIS clusters considered for this study.}
\centering  
\begin{tabular}{ccccc|ccccc}  
\toprule
\toprule
Name & $M_{500}$  & $R_{500}$ & $c$ &$w$ & Name & $M_{500}$  & $R_{500}$ & $c$ &$w$\\
& $10^{14}$ M$_{\odot}$ & Mpc  & & & &$10^{14}$ M$_{\odot}$ & Mpc  & & \\
\midrule
GCL1 &  14.669 & 1.569 & 0.37 & 0.039 & GCL39 &  12.856 & 1.501 & 0.72 & 0.002 \\
GCL2 &  14.993 & 1.580 & 0.30 & 0.036 & GCL40 &  11.167 & 1.432 & 0.39 & 0.034  \\
GCL3 &  13.198 & 1.514 & 0.30 & 0.008 & GCL41 &  10.196 & 1.390 & 0.24 & 0.005  \\
GCL4 &  14.867 & 1.575 & 0.23 & 0.016 & GCL42 &  16.534 & 1.632 & 0.76 & 0.002  \\
GCL5 &  11.301 & 1.438 & 0.39 & 0.027 & GCL43 &  11.983 & 1.466 & 0.30 & 0.034  \\
GCL6 &  11.150 & 1.431 & 0.41 & 0.023 & GCL44 &  9.620 & 1.363 & 0.28 & 0.054 \\
GCL7 &  10.030 & 1.381 & 0.54 & 0.012 & GCL45 &  10.270 & 1.393 & 0.51 & 0.009 \\
GCL8 &  16.231 & 1.622 & 0.76 & 0.002 & GCL46 &  12.541 & 1.489 & 0.82 & 0.005 \\
GCL9 &  13.733 & 1.534 & 0.62 & 0.020 & GCL47 &  9.920 & 1.377 & 0.14 & 0.052 \\
GCL10 &  10.489 & 1.402 & 0.73 & 0.016 & GCL48 &  10.090 & 1.384 & 0.24 & 0.035 \\
GCL11 &  11.480 & 1.446 & 0.34 & 0.004 & GCL49 &  11.768 & 1.457 & 0.27 & 0.015 \\
GCL12 &  12.827 & 1.500 & 0.22 & 0.037 & GCL50 &  9.996 & 1.380 & 0.18 & 0.046 \\
GCL13 &  10.070 & 1.384 & 0.34 & 0.024 & GCL51 &  11.213 & 1.434 & 0.74 & 0.001 \\
GCL14 &  12.475 & 1.486 & 0.85 & 0.007 & GCL52 &  11.262 & 1.436 & 0.11 & 0.039 \\
GCL15 &  14.075 & 1.547 & 0.36 & 0.016 & GCL53 &  9.431 & 1.354 & 0.42 & 0.005 \\
GCL16 &  11.313 & 1.438 & 0.22 & 0.034 & GCL54 &  9.661 & 1.364 & 0.13 & 0.071 \\
GCL17 &  10.496 & 1.403 & 0.47 & 0.004 & GCL55 &  11.096 & 1.429 & 0.30 & 0.025 \\
GCL18 &  12.117 & 1.472 & 0.10 & 0.077 & GCL56 &  15.913 & 1.611 & 0.40 & 0.004 \\
GCL19 &  12.512 & 1.487 & 0.19 & 0.053 & GCL57 &  9.378 & 1.351 & 0.69 & 0.003 \\
GCL20 &  13.004 & 1.507 & 0.15 & 0.010 & GCL58 &  9.782 & 1.370 & 0.29 & 0.005 \\
GCL21 &  11.339 & 1.440 & 0.24 & 0.013 & GCL59 &  10.962 & 1.423 & 0.82 & 0.001 \\
GCL22 &  15.089 & 1.583 & 0.18 & 0.017 & GCL60 &  9.513 & 1.357 & 0.50 & 0.018 \\
GCL23 &  14.979 & 1.579 & 0.31 & 0.006 & GCL61 &  11.893 & 1.463 & 0.36 & 0.032 \\
GCL24 &  15.463 & 1.596 & 0.30 & 0.022 & GCL62 &  9.885 & 1.375 & 0.73 & 0.008 \\
GCL25 &  14.092 & 1.548 & 0.31 & 0.051 & GCL63 &  12.394 & 1.483 & 0.85 & 0.002 \\
GCL26 &  11.478 & 1.446 & 0.32 & 0.022 & GCL64 &  9.678 & 1.366 & 0.68 & 0.002 \\
GCL27 &  15.013 & 1.581 & 0.37 & 0.017 & GCL65 &  10.418 & 1.399 & 0.68 & 0.002 \\
GCL28 &  18.002 & 1.679 & 0.21 & 0.050 & GCL66 &  15.384 & 1.593 & 0.08 & 0.070 \\
GCL29 &  17.014 & 1.648 & 0.70 & 0.007 & GCL67 &  15.476 & 1.596 & 0.10 & 0.065 \\
GCL30 &  12.036 & 1.468 & 0.21 & 0.067 & GCL68 &  10.409 & 1.399 & 0.62 & 0.033 \\
GCL31 &  14.406 & 1.559 & 0.50 & 0.016 & GCL69 &  11.321 & 1.439 & 0.28 & 0.018 \\
GCL32 &  11.797 & 1.458 & 0.41 & 0.022 & GCL70 &  10.038 & 1.382 & 0.24 & 0.015 \\
GCL33 &  14.171 & 1.550 & 0.19 & 0.025 & GCL71 &  11.239 & 1.435 & 0.20 & 0.014 \\
GCL34 &  18.166 & 1.684 & 0.48 & 0.018 & GCL72 &  9.810 & 1.372 & 0.61 & 0.003 \\
GCL35 &  17.227 & 1.654 & 0.25 & 0.025 & GCL73 &  9.382 & 1.351 & 0.38 & 0.035 \\
GCL36 &  14.936 & 1.578 & 0.78 & 0.001 & GCL74 &  9.936 & 1.378 & 0.24 & 0.012 \\
GCL37 &  13.217 & 1.515 & 0.28 & 0.045 & GCL75 &  10.774 & 1.415 & 0.42 & 0.032 \\
GCL38 &  9.899 & 1.375 & 0.32 & 0.042 & & & & & \\ 
\bottomrule                               
\end{tabular}
\tablefoot{In the Table, we list $M_{500}$, $R_{500}$, and the morphological parameters $c$ and $w$, computed as described in \citet{towler23}.}             
\label{table:MACSIS}
\end{table*}

\begin{figure*}
            \centering 
            \includegraphics[width=0.40\textwidth]{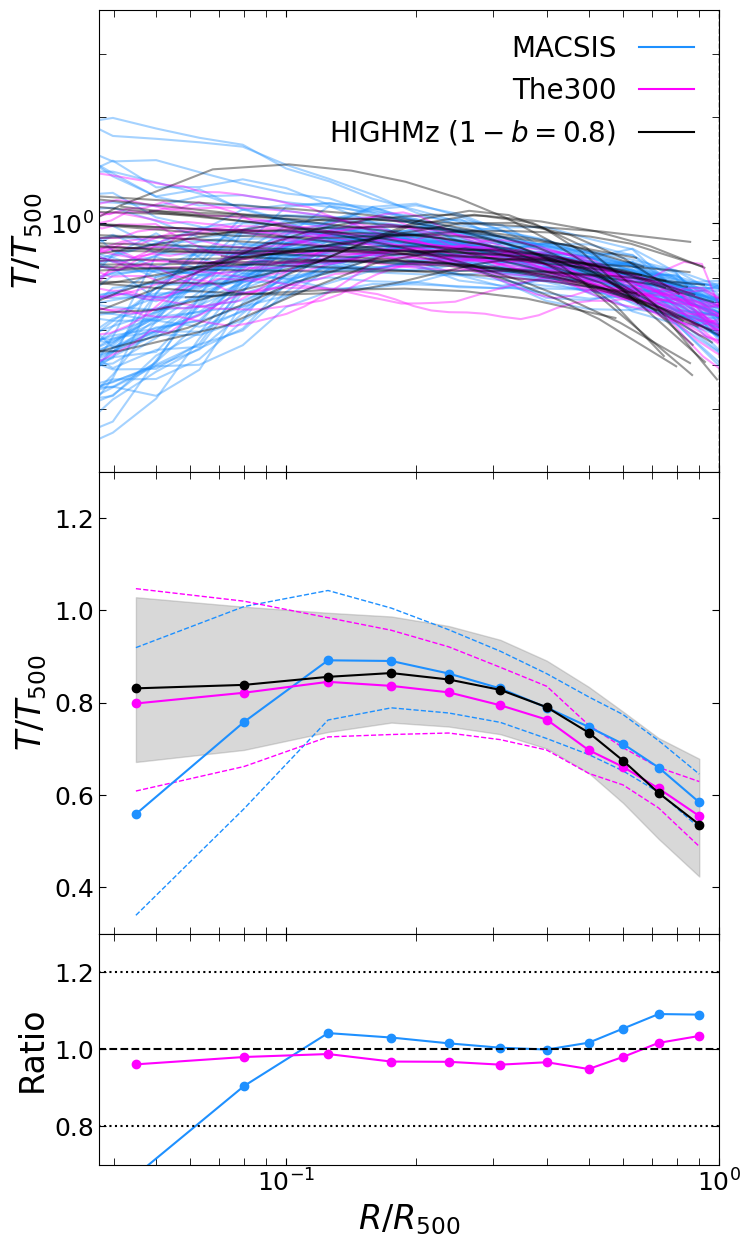}
            \includegraphics[width=0.41\textwidth]{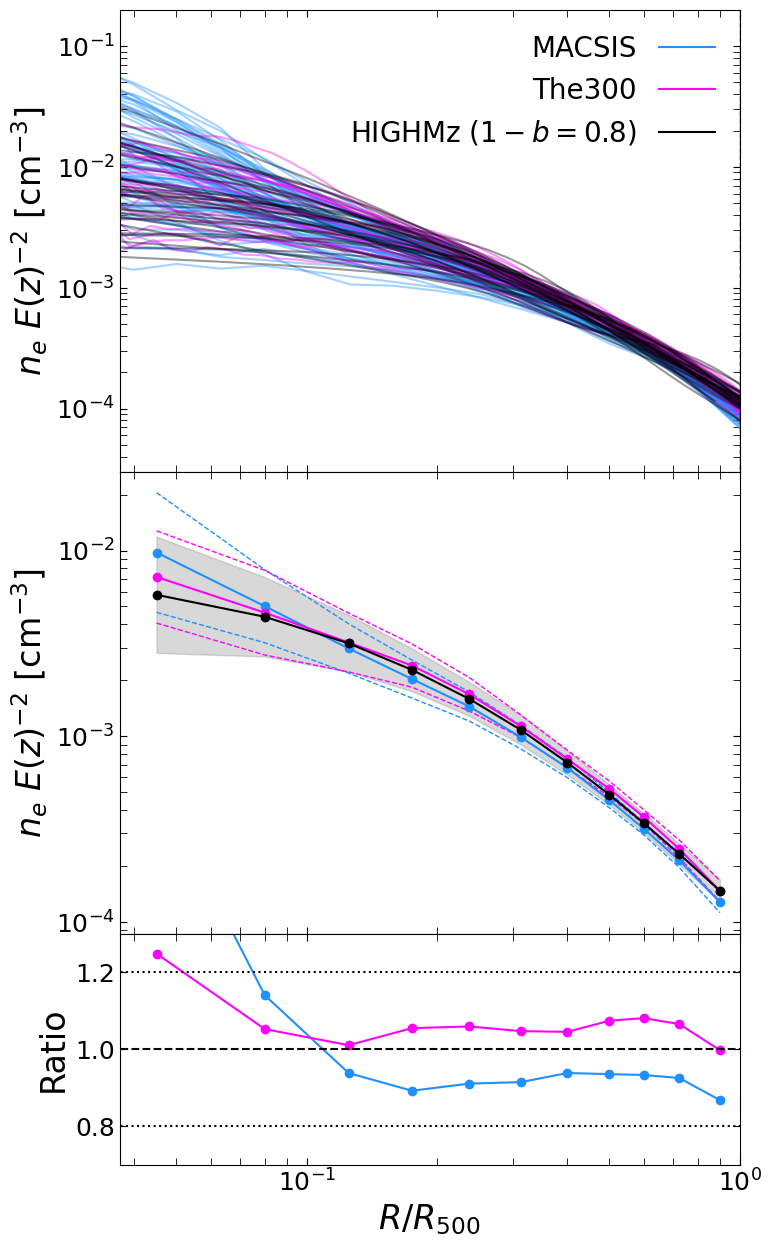}
            \caption{\footnotesize Comparison of temperature (left) and density (right) profiles of MACSIS (light blue) and The300 (violet) clusters, with respect to HIGHMz ones (black). Upper, central and lower panels show the same information as in Fig. \ref{K_sim}.}
            \label{T_ne_sim}
\end{figure*}

In Sect. \ref{simulations}, we showed that the scaled entropy profiles of HIGHMz clusters lie between those of The300 and MACSIS clusters, in terms of their normalisation, albeit with some small systematic differences at all radii. To better understand the origin of these discrepancies, and in general to further investigate the comparison with the simulated datasets, we broke down the contributions of MACSIS and The300 temperature and electron density profiles; these profiles are shown in Fig. \ref{T_ne_sim}, in comparison with HIGHMz. In order to compute the median and the intrinsic scatter of the profiles, we followed the same procedure outlined in Sect. \ref{sec_ent_results}. The temperature profiles were scaled using $T_{500}$ computed from Eq. 2 of \citet{arnaud05}, for an overdensity $\Delta=500$ and $T > 3.5$ keV; densities are scaled self-similarly by $E(z)^2$.

We notice that, beyond the core, the median temperature profile of the The300 systematically underestimates the observed one, while densities are overestimated. A similar, but opposite, behaviour is found for MACSIS. Given the way temperature and density combine to yield the entropy, the resulting entropies are underestimated for The300 and overestimated for MACSIS, as already observed in Fig. \ref{K_sim}. When looking at the central regions ($\lesssim 0.1~R_{500}$), the differences between simulations and observations become more evident. MACSIS clusters exhibit, in median, a stronger cooling and a more peaked central density profile than observed in HIGHMz clusters, as also found in previous works comparing simulations to observations (\citeauthor{barnes17} \citeyear{barnes17}; \citeauthor{braspenning24} \citeyear{braspenning24}), while The300 clusters are better at reproducing the central temperature distribution of observations. Once again, these discrepancies are the origin of the differences noted in Fig. \ref{K_sim} for the entropy profiles.
\begin{figure}
\centering
            \includegraphics[width=0.48\textwidth]{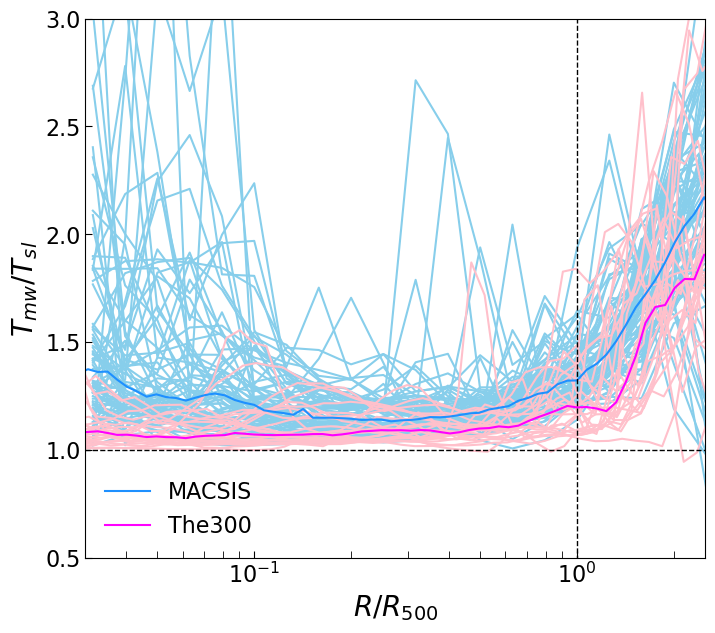}
            \caption{\footnotesize Comparison of mass-weighted ($T_{mw}$) and spectroscopic-like ($T_{sl}$) temperature profiles, for MACSIS (light blue) and The300 (pink). Plotted are the ratios of the two measurements for each cluster. The two more marked lines are the median ratios for the two samples. }
            \label{Tmw_sl_sim}
\end{figure}

As a final consideration, we recall that in Sect. \ref{simulations} we considered mass-weighted entropy profiles for the simulated clusters, in comparison with HIGHMz ones. \citet{mazzotta04} discussed an alternative weighting method of the particles in simulations, called spectroscopic-like, which weights more dense and cold sub-structures, as it may happen for real observations if these regions are not masked properly in the analysis. However, since sub-structures are not masked in simulations, the use of spectroscopic-like profiles may bias the reproduced temperatures towards excessively low values. For this reason, in Sect. \ref{simulations}, we preferred to keep mass-weighted profiles as our reference. However, we present in Fig. \ref{Tmw_sl_sim} a direct comparison of mass-weighted to spectroscopic-like temperature profiles. In general, spectroscopic-like temperature profiles are lower than mass-weighted ones at all radii, particularly so in the cluster outskirts, where the presence of in-falling sub-structures causes a net increase of the ratio $T_{mw}/T_{sl}$. Another consequence of that is also the increase of the intrinsic scatter for spectroscopic-like profiles, which becomes consistent with observations at $\sim R_{500}$, both for MACSIS and The300.

\section{Residual dependence on redshift}
\label{App_Az_dependence}
\begin{figure}
            \centering
            \hspace{-0.2cm}\includegraphics[width=0.45\textwidth]{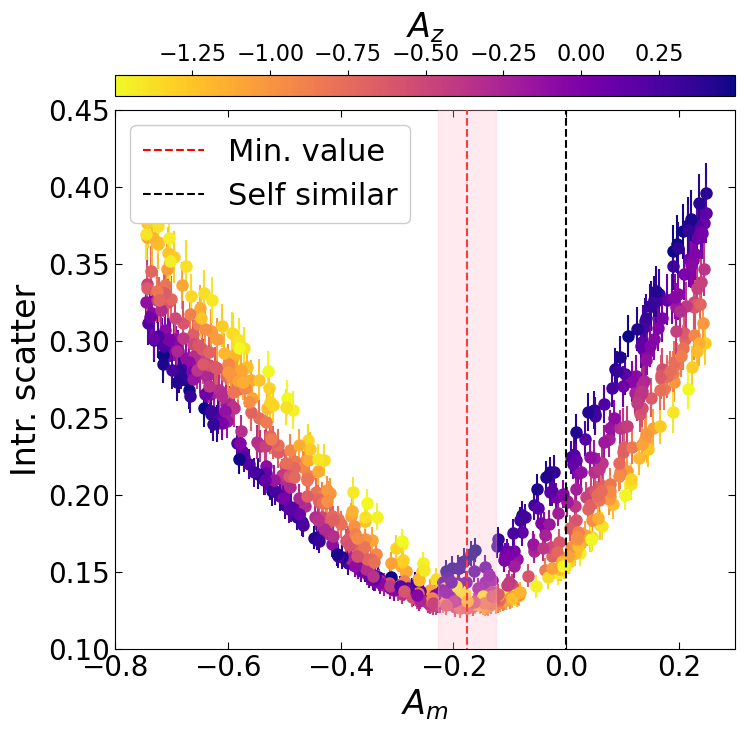}
            \includegraphics[width=0.46\textwidth]{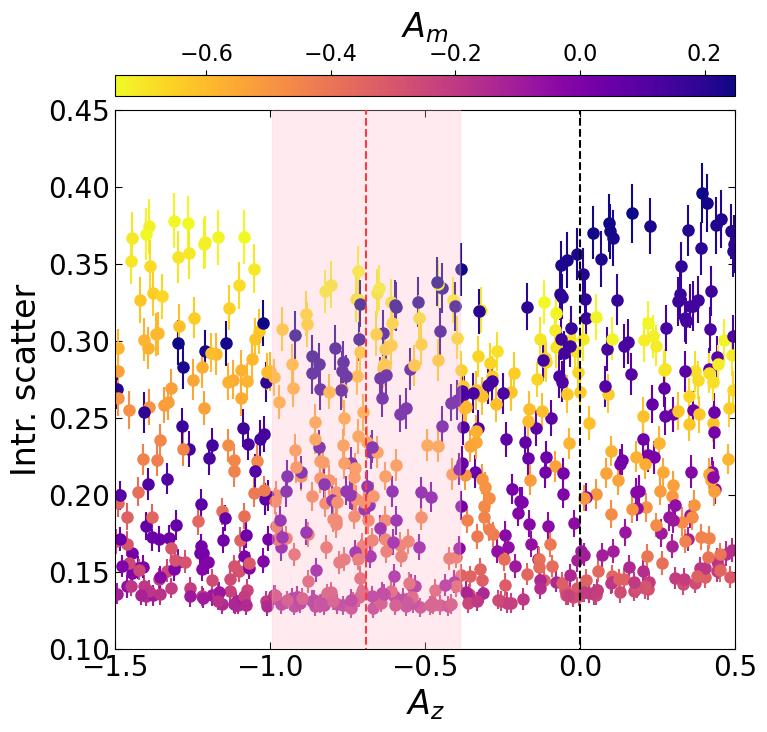}
            \caption{\footnotesize Constraints to both residual mass ($A_m$) and redshift ($A_z$) dependencies, as defined in Eq. \ref{eq_az}. In the figure, we plot the distribution of the intrinsic scatter of the combined HIGHMz -- \rexcess\ -- ESZ sample, as a function of $A_m$ (top) and $A_z$ (bottom). In the upper panel, measured points are colour coded according $A_z$, while in the lower one according to $A_m$ values. Vertical red lines mark the positions of the two minimum values, while black ones are predictions from the self-similar scenario (i.e. $A_m=A_z=0$).}
            \label{Az_surface}
\end{figure}

We further explore the possibility of an alternative rescaling of the entropy profiles that goes beyond self similarity (Eq. \ref{eq_k500}). In particular, in this section we investigate a possible residual dependence on both mass and redshift, using the clusters in the combined HIGHMz -- \rexcess\ -- ESZ sample. Similarly to Sect \ref{Am_section}, we introduced two parameters, $A_m$ and $A_z$, which quantify the departure from self-similar predictions for the mass and redshift dependencies, respectively, and built the modified entropy rescaling:
\begin{equation}
    \widetilde{K_{500}} = M_{500}^{2/3 + A_m}~E(z)^{-2/3+A_z}.
    \label{eq_az}
\end{equation}
We generated $500$ random values of $A_m$ and $A_z$, uniformly distributed in the ranges $[-0.75,+0.25]$ and $[-2.5,+1.0]$, respectively. At each iteration, we rescaled the entropy profiles using a different realisation of $\widetilde{K_{500}}$ and fitted them in the radial range $[0.3-0.8]~R_{500}$, using the model described in Eq. \ref{pow_eq}. We then studied how the measured scatter varies as a function of $A_m$ and $A_z$; our best-fitting residual mass and redshift dependencies are the values of $A_m$ and $A_z$ that minimise the dispersion of the profiles. The results of this procedure are shown in Fig. \ref{Az_surface}.

The distribution of the intrinsic scatter $\sigma_{\mathrm{int}}$ as a function of $A_m$ and $A_z$ describes a well defined surface in the 3D parameter space and the minimum of it is not achieved for values $(A_m,A_z)=(0,0)$, as it would be the case if entropy profiles scaled self similarly. The projection of the 3D distribution along the $A_m$ (top) and $A_z$ (bottom) axes allows for a more precise identification of the two values for which the measured scatter is at its minimum. In particular, we find that the pair $(A_m,A_z)=(-0.18 \pm 0.05,-0.69 \pm 0.30)$ is able to minimise the dispersion of the considered entropy profiles. The dependence of the intrinsic scatter on $A_m$ is stronger, as suggested by the well defined minimum and by the evident curvature, while the one on $A_z$ is less pronounced; this is likely simply due to the redshift range of the considered clusters ($z < 0.6$), which does not allow more robust constraints on this parameter. The measured value $A_m$ that minimises the dispersion is in agreement with predictions by \citet{ettori23}, who found $A_m= -0.16 \pm 0.01$; however, they also derived a weaker residual dependence on redshift ($A_z = -0.07 \pm 0.02$) than our measurement, although this is a $2\sigma$ effect. Similar results were found by \citet{pratt22} regarding the density profiles of $\sim 120$ clusters up to $z\sim1.1$. If we assume that the departure from self-similar predictions comes from density profiles only, their measurements would translate into $A_m \sim -0.15$ and $A_z \sim -0.06$ for the ICM entropy. Finally, given the adopted colour coding, we notice that, at fixed $A_z$ (top) and $A_m$ (bottom), the distribution of $\sigma _{int}$ describes well defined curves in the parameter space; for example, for $A_z=0$ we obtain the result of Fig. \ref{Am}.

\section{ICM entropy from mean density and other mass rescalings}
\label{sect_mean_entropy}

\begin{figure}
\centering
            \includegraphics[width=0.5\textwidth]{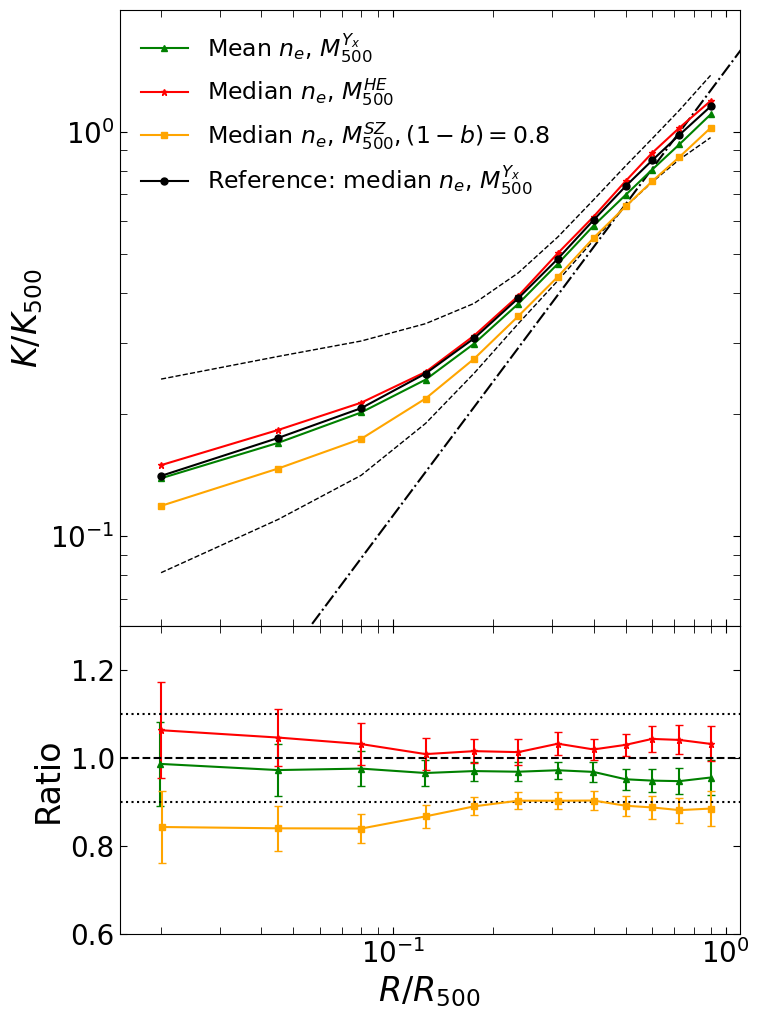}
            \caption{\footnotesize Different rescalings presented throughout the paper. In particular, we compare profiles derived with i) mean densities, scaled by $M_{500}^{Y_{\rm X}}$ (green triangles); ii) median densities, scaled by $M_{500}^{\rm HE}$ (red stars); iii) median densities, scaled by $M_{500}^{Y_{\rm SZ}}$ ($1-b=0.8$; orange squares) to the reference ones, i.e. obtained with median densities and scaled by $M_{500}^{Y_{\rm X}}$ (black dots).}
            \label{scalings}
\end{figure}
\begin{figure*}
            \centering
            \hspace{0.245cm}\includegraphics[width=0.837\textwidth]{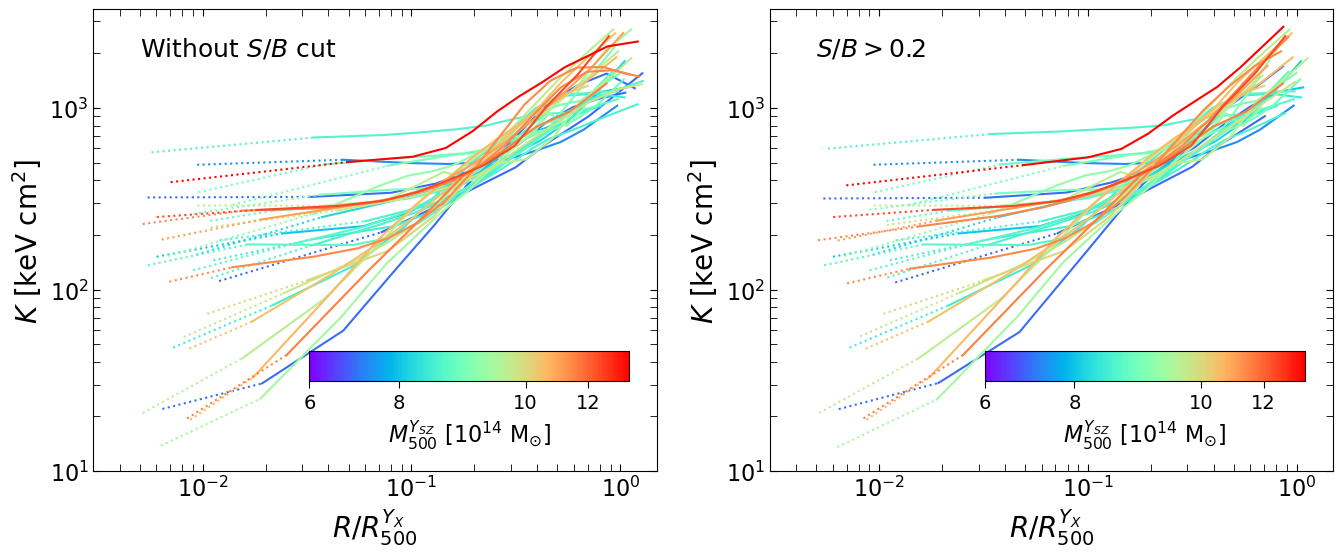}
            \includegraphics[width=0.85\textwidth]{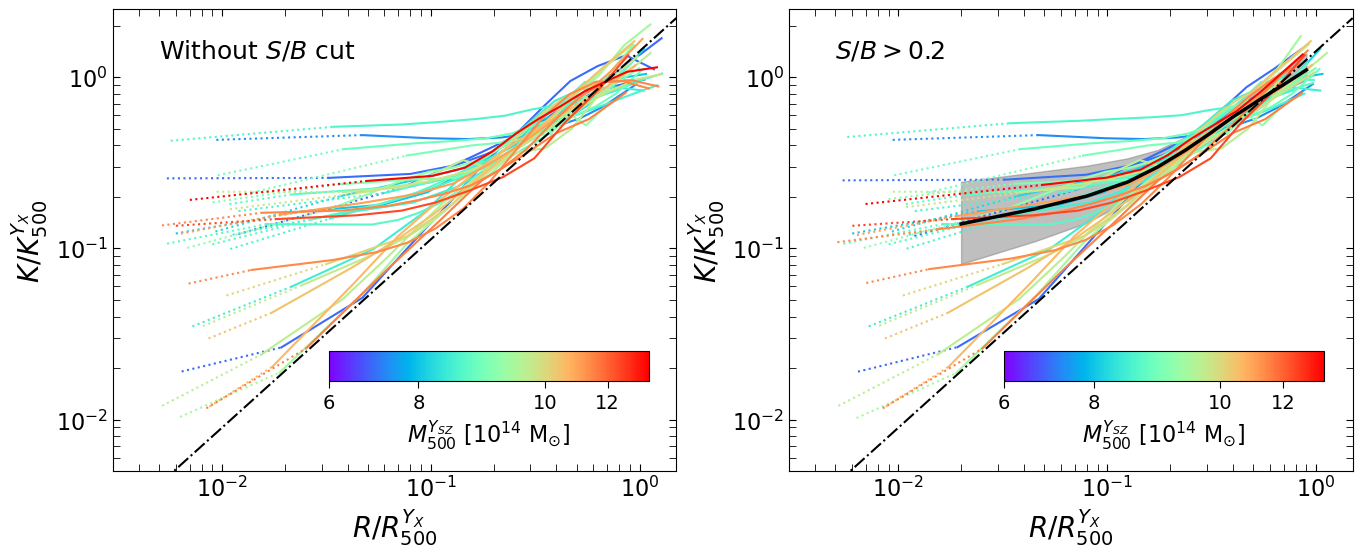}
            \includegraphics[width=0.85\textwidth]{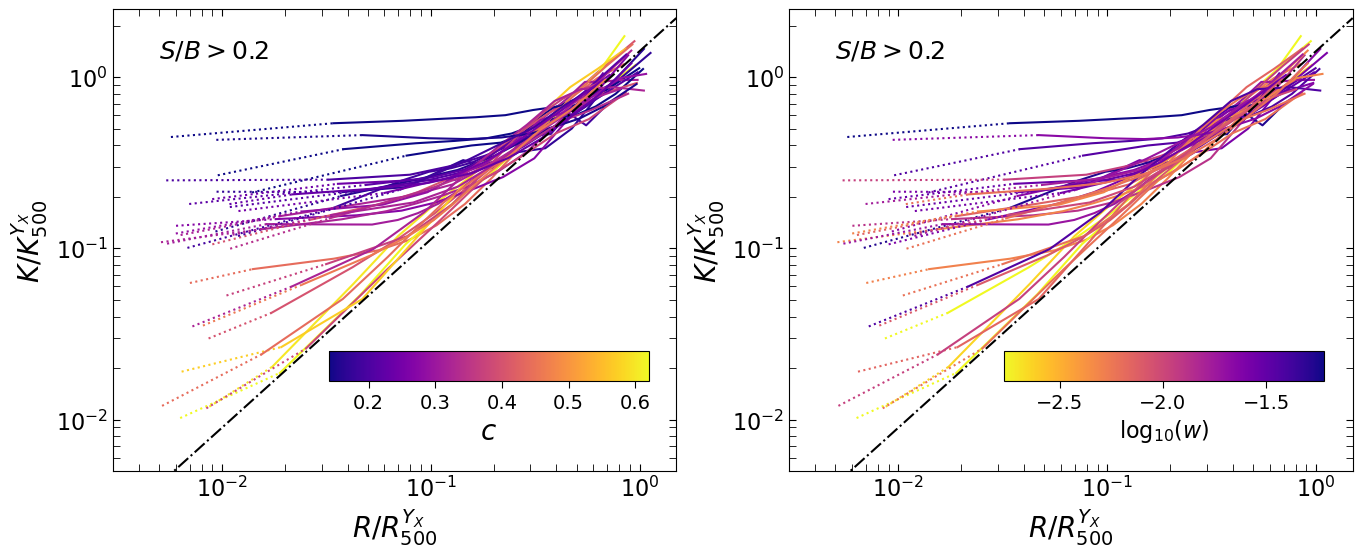}
            \caption{\footnotesize Entropy profiles of HIGHMz clusters, derived from azimuthal mean densities. The description to the figure is the same as for for Fig. \ref{K_prof}.}
            \label{K_prof_mean}
\end{figure*}

Although we considered the profiles obtained from median densities and scaled by $M_{500}^{Y_{\rm X}}$ as our reference, a combination of other entropy measurements (e.g. from mean densities) and mass rescalings (e.g. by $M_{500}^{\rm HE}$ and $M_{500}^{Y_{\rm SZ}}$, with $1-b = 0.8$) were used throughout the paper for comparison with other samples. In this section, we show the differences of these scaled profiles with respect to the reference ones (Fig. \ref{scalings}), with particular focus on those derived from azimuthal mean densities and scaled by $M_{500}^{Y_{\rm X}}$ (Fig. \ref{K_prof_mean}), which were used in comparison with \rexcess\ and ESZ samples in Sect. \ref{other_samples}. 

In general, the profiles shown in Fig. \ref{K_prof_mean} have similar characteristics to those presented in Sect. \ref{sec_ent_results}, such as the observed large dispersion in the central regions, which reflects the morphological state of the clusters, and a smaller one towards the outskirts. Also here, the exclusion of temperature bins with $S/B < 0.2$ from the deprojection allows a regularisation of the profiles at all radii and alleviates the observed flattening at $\sim R_{500}$. Conversely, using mean densities, which have better resolution than median ones, we can explore further the core region of the clusters and derive the ICM entropy (assuming a constant core temperature) down to $R \sim 0.005~R_{500}^{Y_{\rm X}}$. A quantitative comparison between azimuthal median and mean scaled entropy profiles (Fig. \ref{scalings}, green triangles) shows that the latter are $\lesssim 5\%$ lower at all radii, a small effect given both to higher densities and masses derived from the $Y_X$ proxy, adopted in the scaling.

To compare with other samples, both observational and simulated ones, we also used two additional different versions of the scaled entropy profiles. In particular, we used profiles derived from median densities and scaled by $M_{500}^{\rm HE}$ in comparison with X-COP (Sect. \ref{other_samples}) and $M_{500}^{Y_{\rm SZ}}$ masses, corrected to mimic a hydrostatic bias of $1-b=0.8$, in comparison with MACSIS and The300 (Sect. \ref{simulations}). The medians of these scaled profiles are shown in Fig. \ref{scalings} (red stars and orange squares, respectively), compared to the reference one. Using hydrostatic-equilibrium masses to compute $K_{500}$ (Eq. \ref{eq_k500}) leads to a small increase ($\lesssim 3\%$) of the scaled entropy, not significant though; correcting \textit{Planck} masses and subsequently using them in the rescaling makes the scaled entropy a $\sim 10\%$ lower with respect to the reference profile.


\end{appendix}

\end{document}